\documentclass[%
reprint,
 amsmath,amssymb,
prx,
]{revtex4-2}
\allowdisplaybreaks

\newcommand{\appsection}[1]{\section{\MakeUppercase{#1}}}

\usepackage{graphicx}
\usepackage{dcolumn}
\usepackage{bm}
\usepackage{xcolor,soul}

\usepackage{hyperref}
\hypersetup{
  colorlinks=false,
  linktoc=all
}
\usepackage[capitalise]{cleveref}

\DeclareMathOperator{\tr}{tr}
\newcommand{\rmd}{\mathrm{d}}
\newcommand{\sech}{\mathrm{sech}}

\begin{document}

\preprint{APS/123-QED}

\title{Topological Quantum Error-Correcting Thresholds \\of a Symmetry-Enriched Toric Code}

\author{Jiasheng He}
 \email{jshhe@physics.ucla.edu}
\author{Ryan Thorngren}
 \email{ryan.thorngren@physics.ucla.edu}
\affiliation{Mani L. Bhaumik Institute for Theoretical Physics, Department of Physics and Astronomy,\\
University of California, Los Angeles, California 90095, USA}

\date{\today}

\begin{abstract}
We study the error thresholds of a new topological quantum error-correcting code under the independent bit-flip and phase-flip noisy channel. By mapping to classical spin-glass models, we find that if Pauli $X$ and $Z$ errors are equally probable, the threshold of vertex syndromes is improved by 30\% compared to that of the toric code,  while the threshold of plaquette syndromes is lowered. Both thresholds increase when the $Z$ component is less likely than the $X$ component, reaching 0.5 if the $Z$ component is absent. An analytical proof justifies this special case of perfect correctability, and a disordered version of Peierls's argument establishes a lower bound for the vertex-syndrome threshold.
\end{abstract}

\maketitle


\section{Introduction}
Quantum error correction (QEC) \cite{Shor1995,Steane1996,Bennett1996,Knill1997} is indispensable for building quantum computers because of the extreme sensitivity of qubits to environmental noise.  In recent years, topological error-correcting codes have drawn significant interest since they encode the logical information nonlocally and thus protect against local physical errors. A famous example is Kitaev’s toric code \cite{Kitaev2003} where physical qubits are designed on a 2-torus, and the logical qubits are encoded as the parity of the first homology group of the 2-torus. For example, the logical state $|01\rangle$ stands for the state where the physical qubits represent the homology class of a meridian of the torus. To prevent the accumulation of local errors into nonlocal, logical errors in time, repeated syndrome measurements and corrections are necessary for the proper memory of a topological quantum computer \cite{Dennis2002,Fowler2012}. Nevertheless, this kind of topological protection is only feasible to a certain degree. If the physical error rate exceeds a threshold \cite{Knill1998}, the complexity of error syndromes will lead to ambiguous and usually wrong error correction. For example, if all but two physical qubits on a whole meridian suffered errors, the error recovery could incorrectly fix the two to cure the syndrome but destroy the stored memory. Therefore, inventing a topological code that is robust to a high error rate becomes a crucial and promising resolution to the problem. 

\begin{figure}[b]
\centering
\includegraphics[width=\columnwidth]{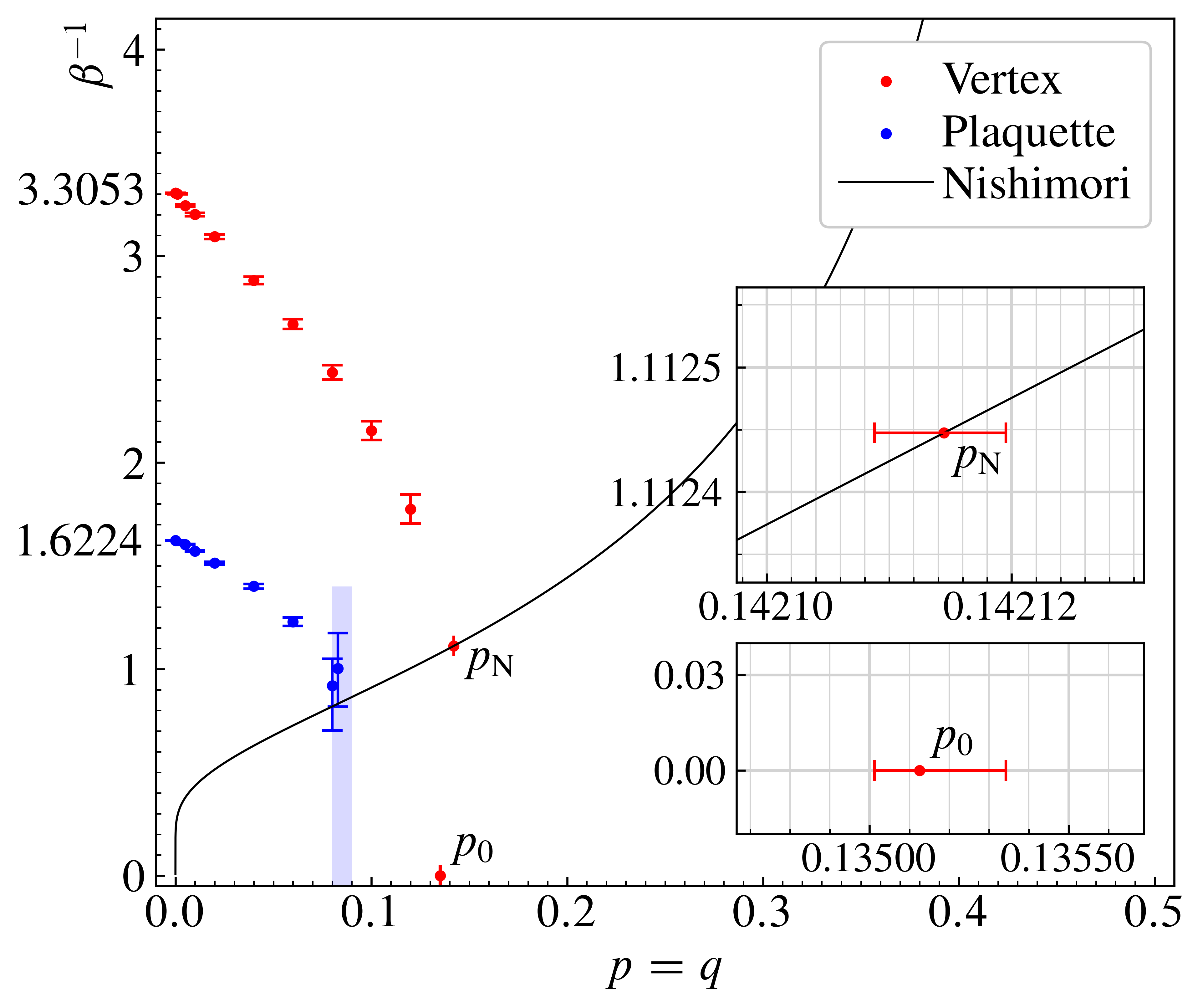}
\caption{\label{fig:phasediagram_pqsame} Phase diagram in the $p=q$ plane of the SM models for decoding vertex syndromes, \cref{eq:SM_Hamiltonian}, and plaquette syndromes, \cref{eq:SM_Hamiltonian_dual}. Every point below the critical temperatures is in the ferromagnetic phase, while the rest is in the paramagnetic phase. The Nishimori point of vertex syndromes is $p_\mathrm{N} = 0.14211448$ with a 95\% confidence interval $(0.14210878, 0.14211952)$, while that of plaquette syndromes is between 0.083 and 0.09.} 
\end{figure}

The study of the error threshold of a topological code can be transformed into the study of the phase transition of a corresponding classical, disordered statistical mechanical (SM) model \cite{Dennis2002,Nishimori,WANG200331,Raussendorf2007,Katzgraber2009,Bombin2010,Chubb2021}. Kitaev’s toric code has been mapped to the random-bond Ising model \cite{Dennis2002}, under the assumption that Pauli $X$ and $Z$ errors are equally probable but independent, and the error threshold is found to be about $0.1094 \pm 0.0002$, which is the Nishimori point $p_\mathrm{N}$, i.e., the intersection point of the ferro-paramagnetic phase boundary and the Nishimori line of the random-bond Ising model \cite{Honecker2001,Picco2006}.

\begin{figure}[h]
\centering
\includegraphics[width=\columnwidth]{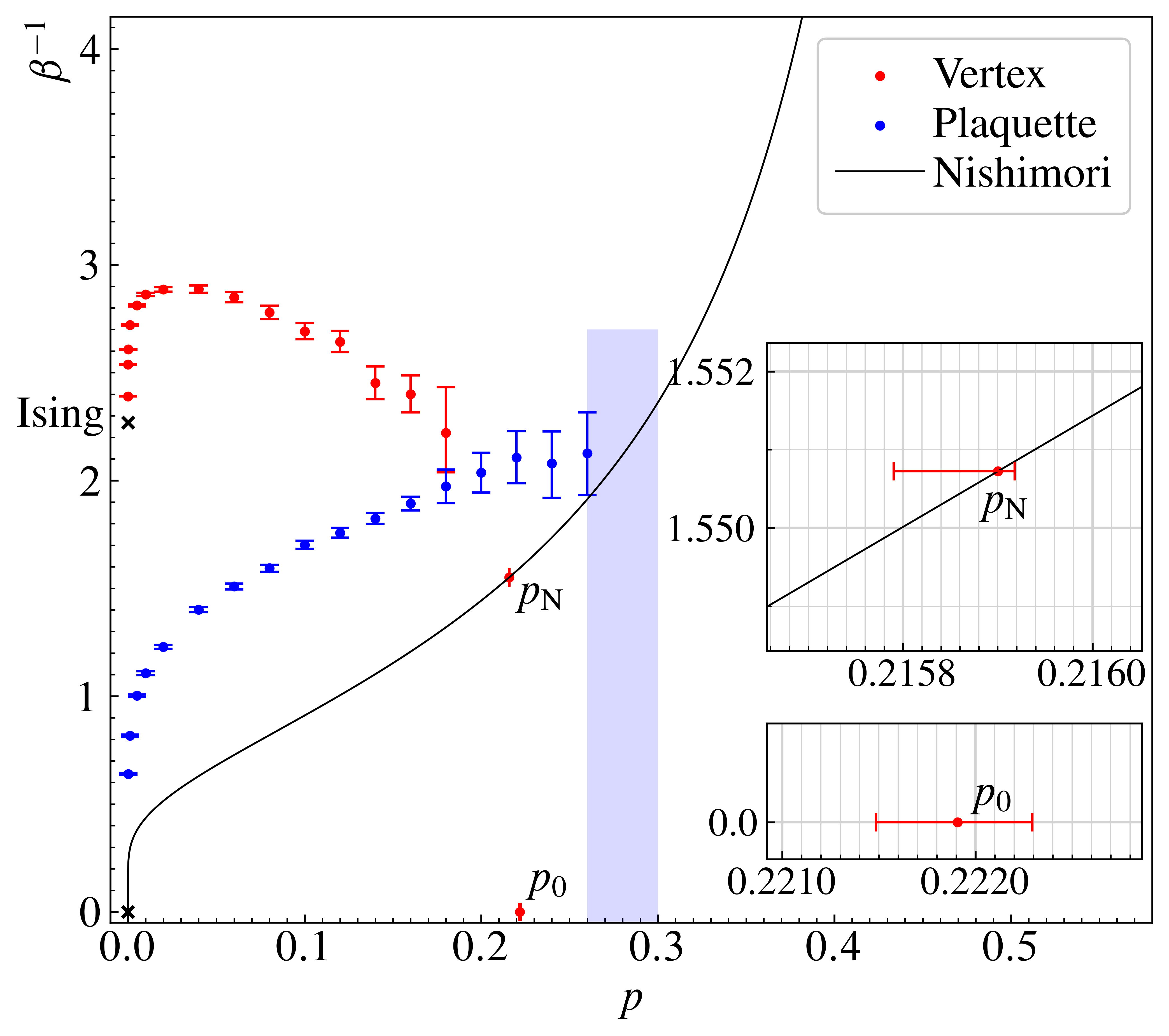}
\caption{Phase diagram in the $q=0.04$ plane of the SM models for decoding vertex syndromes, \cref{eq:SM_Hamiltonian}, and  plaquette syndromes, \cref{eq:SM_Hamiltonian_dual}. The Nishimori point of vertex syndromes is $p_N = 0.21590018$ with a 95\% confidence interval $(0.21579023,0.21591802)$, while that of plaquette syndromes is between 0.26 and 0.3. ``Ising'' stands for the Ising model's critical temperature $2/\ln(1+\sqrt{2})$.} \label{fig:phasediagram_q04}
\end{figure}

In this paper, we study a model closely related to the toric code, which was introduced in \cite{ohyama2026parameterizedfamiliestoriccode} in the context of topological families. This new code displays an enhanced error threshold for vertex syndromes via a novel geometric mechanism. To understand this mechanism, consider the correspondence between toric-code error correction and the random-bond Ising model. In that context, two error chains that produce the same syndrome can always be deformed into each other by a closed loop, and this closed loop is thought of as a domain wall (DW) of the Ising model. The random bond is then due to the fact that we cannot exactly detect where the errors are, so any supposed error configuration for a measured syndrome is then thought of as a disordered bond realization. If we know more about the error chains, the possible types of deformation can be limited, and the resulting error threshold will be improved. This is the motivation for this code. For vertex syndromes, while the toric code detects only the endpoints, this new code allows measurement of the corners as well as the endpoints of an error chain. The error threshold is 
\begin{align}
p_\mathrm{N} = 0.14211448 \label{eq:p_N}
\end{align}
with a 95\% confidence interval 
\begin{align}
(0.14210878,\, 0.14211952).
\end{align}
This is a 30\% improvement over that of the toric code under the same conditions. For plaquette syndromes, the corresponding SM model is dual to the other and thus is expected to have a threshold that is lower than that of the toric code. We found
\begin{align}
0.083< p'_\mathrm{N} < 0.09.
\end{align}
Note that the toric code is self-dual and thus has the same threshold for vertex and plaquette syndromes. Under the condition that $Z$ is less likely than $X$, both thresholds can be further improved and even reach 0.5 in the case of no $Z$ component \footnote{Note that we only care about physical error rates below 0.5. Otherwise, no encoding can preserve arbitrary unknown quantum states. At 0.5, every physical qubit is completely random. Above 0.5, more than half of the information is lost to the environment; if we could recover the state of the logical qubits, the environment would also be able to recover it, but this is impossible according to the no-cloning theorem}. See \cref{fig:phasediagram_pqsame,fig:phasediagram_q04}.

\section{Description of the QEC code}
Consider a square lattice with periodic boundary conditions in both directions. Unlike the toric code which only has qubits on links, this new code has qubits on both links and vertices. The Hamiltonian is 
\begin{align}
H = -\sum_i R_i - \sum_i V_i - \sum_p B_p, \label{eq:qec_Hamiltonian}
\end{align}
where the sums are over all vertices $i$ and plaquettes $p$, and the operators $R_i$, $V_i$, and $B_p$ are tensor products of Pauli operators on the following local qubits:
\begin{align}
R_i &= \vcenter{\hbox{\includegraphics{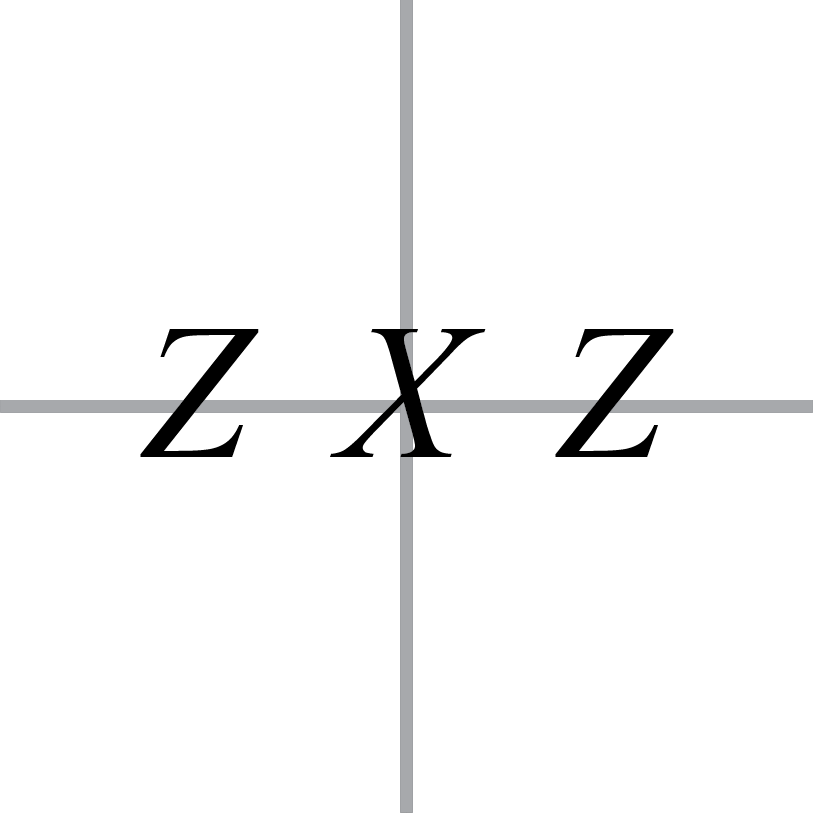}}},\\
V_i &= \vcenter{\hbox{\includegraphics{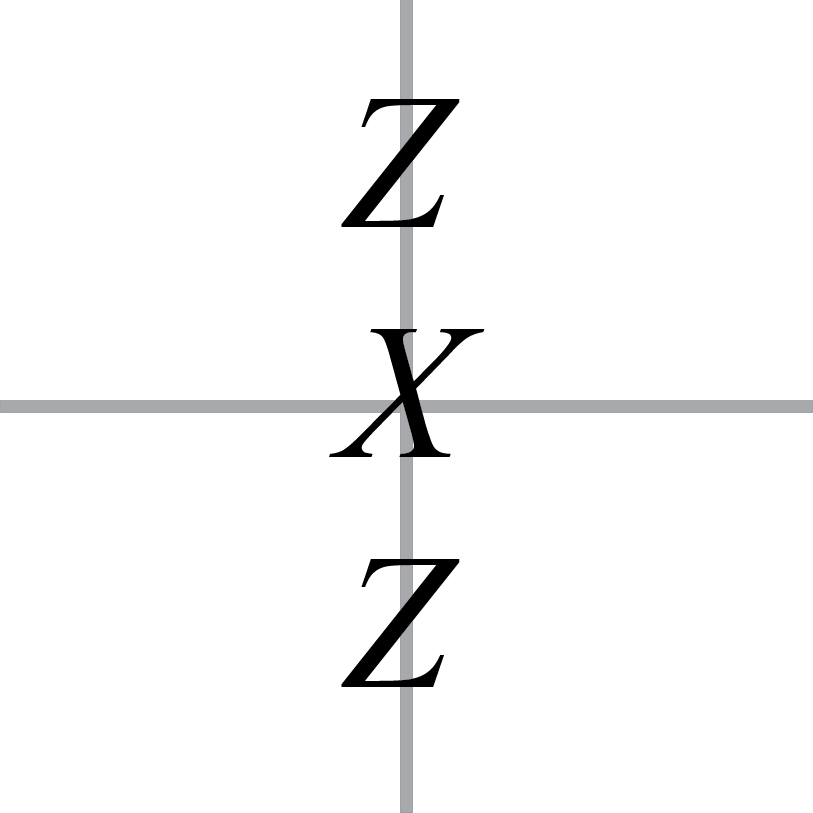}}}, \\
B_p &= \vcenter{\hbox{\includegraphics{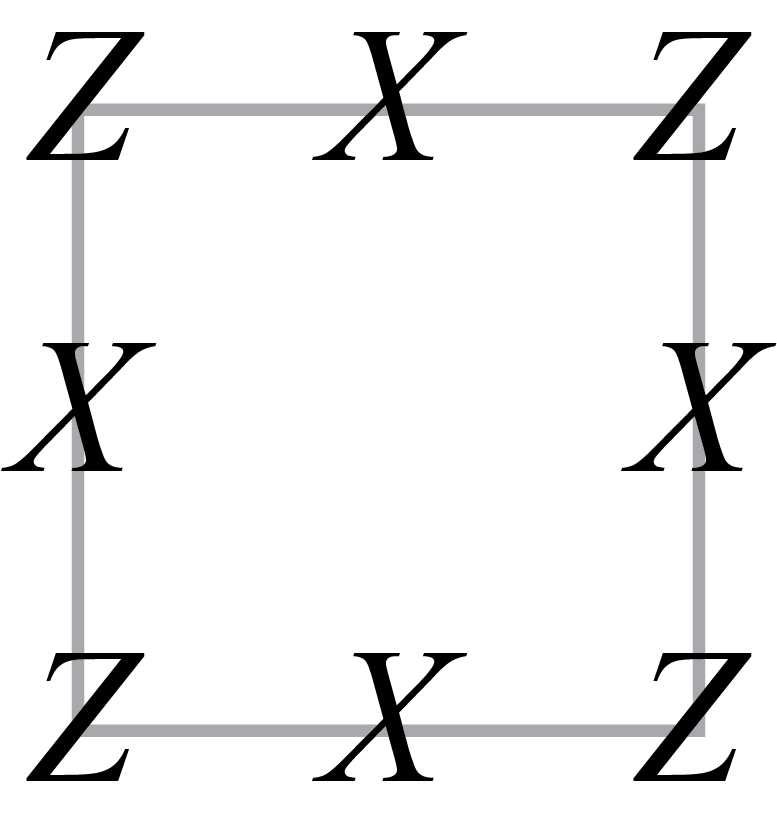}}}.
\end{align}
The eigenstates of $Z$ are called the up state $\left|\uparrow\right\rangle$ and the down state $\left|\downarrow\right\rangle$ and satisfy $Z\left|\uparrow\right\rangle = \left|\uparrow\right\rangle$ and $Z\left|\downarrow\right\rangle = -\left|\downarrow\right\rangle$, while the eigenstates of $X$ are $|\pm\rangle = (\left|\uparrow\right\rangle \pm \left|\downarrow\right\rangle)/\sqrt{2}$ and satisfy $X|\pm\rangle = \pm |\pm\rangle$.

We will see below that this stabilizer Hamiltonian has ground states related by a finite-depth quantum circuit to the ground states of the usual toric code. Note that it can be put in the Calderbank-Shor-Steane (CSS) form by a Hadamard transformation on all vertices.

\subsection{Ground states}
The ground states of our Hamiltonian \cref{eq:qec_Hamiltonian} satisfy 
$R_i |\text{GS}\rangle = + |\text{GS}\rangle$, $V_i |\text{GS}\rangle = + |\text{GS}\rangle$, and $B_p|\text{GS}\rangle = + |\text{GS}\rangle$ for every vertex $i$ and plaquette $p$. To find them, we first define the following operator for every vertex $i$: 
\begin{align}
C_i = \exp \left( \mathrm{i}\pi \frac{1-Z_i}{2}\otimes \frac{1-Z_{\text{left}(i)}\otimes Z_{\text{right}(i)}}{2} \right) ,
\label{eq:qec_C}
\end{align}
where left($i$) is the link emanating from the vertex $i$ to the left, and right($i$) is the link emanating from the vertex $i$ to the right, and $\mathrm{i}\equiv\sqrt{-1}$. By using the anticommutation relations of Pauli operators, we can show that 
\begin{subequations}
\begin{align}
C_i \ \vcenter{\hbox{\includegraphics{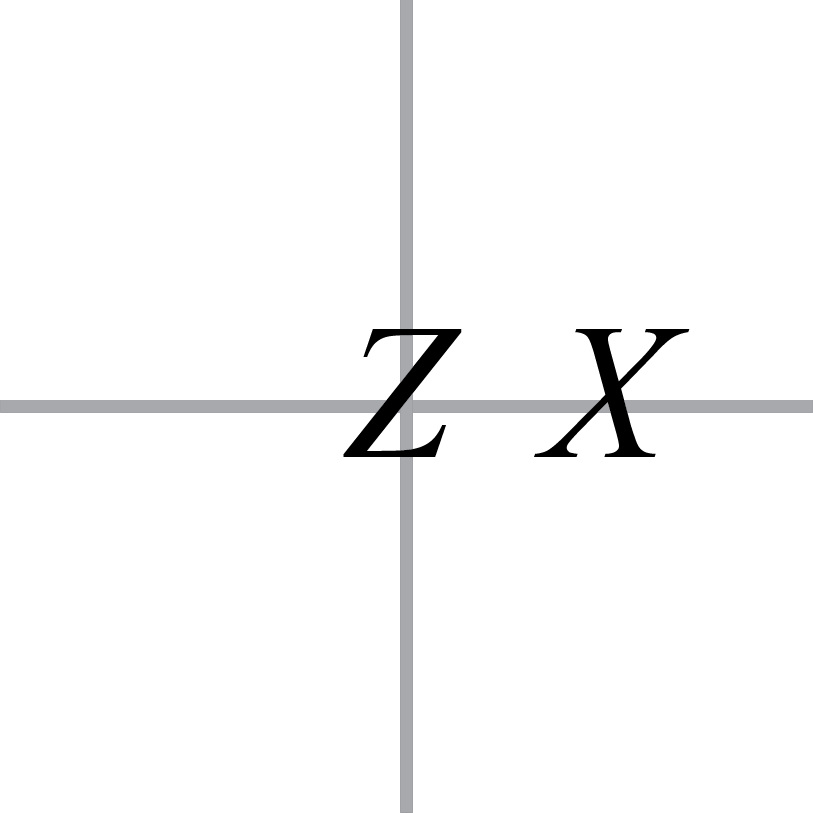}}}\ C_i \ &= \ \vcenter{\hbox{\includegraphics{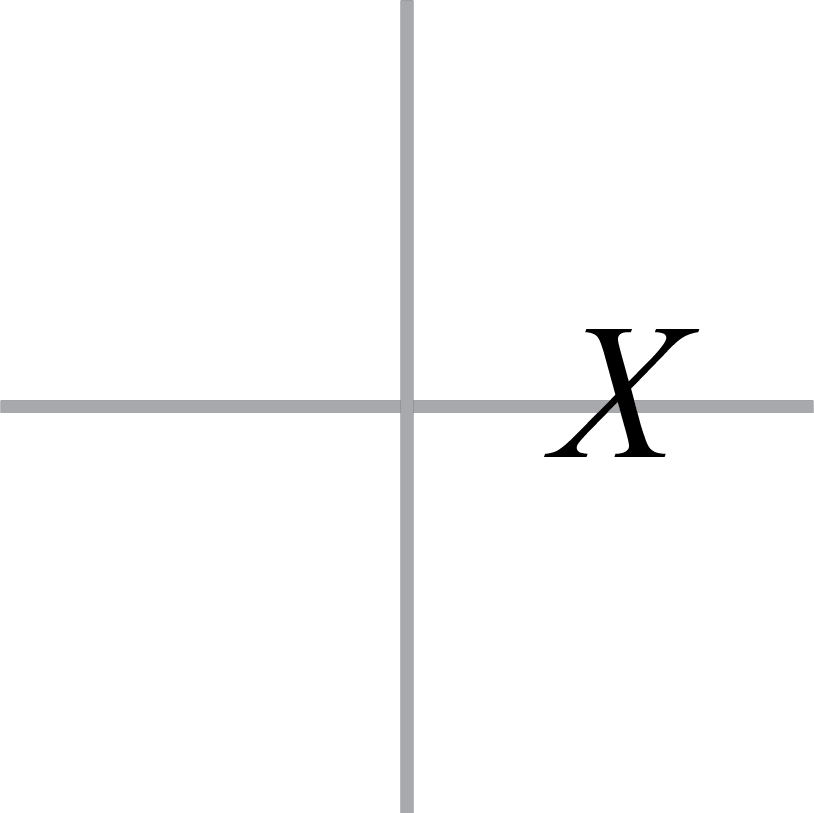}}},\\
C_i R_i C_i \ =\  C_i \ \vcenter{\hbox{\includegraphics{R.png}}}\ C_i \ &= \ \vcenter{\hbox{\includegraphics{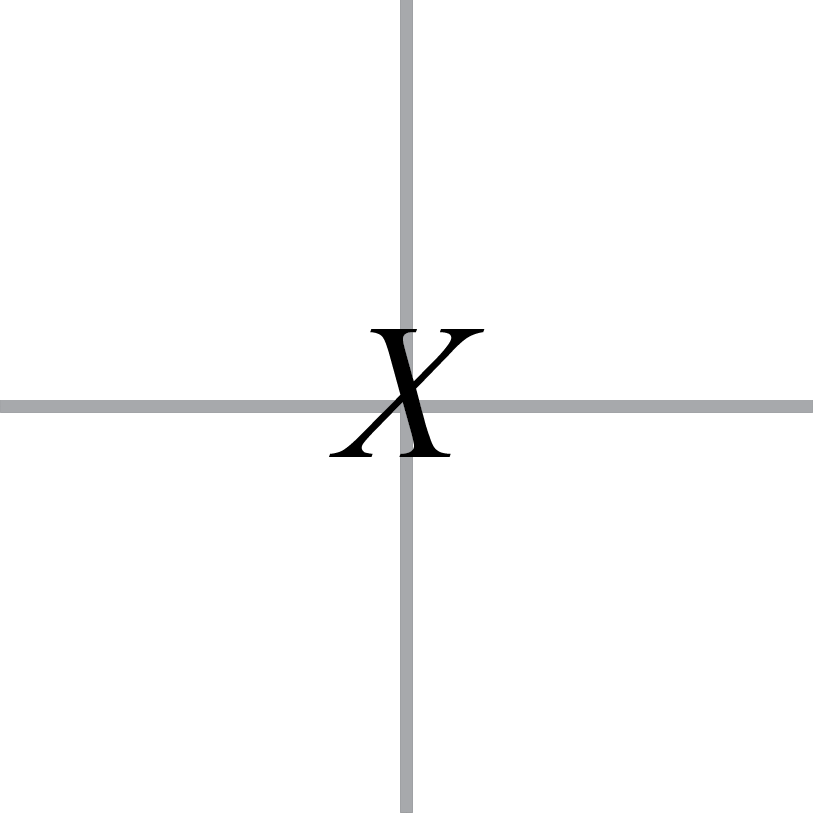}}}, \\
C_i \ \vcenter{\hbox{\includegraphics{IXI.png}}}\ C_i \ &= \ \vcenter{\hbox{\includegraphics{R.png}}},\\
C_i V_i C_i \ =\  C_i \ \vcenter{\hbox{\includegraphics{V.png}}}\ C_i \ &= \ \vcenter{\hbox{\includegraphics{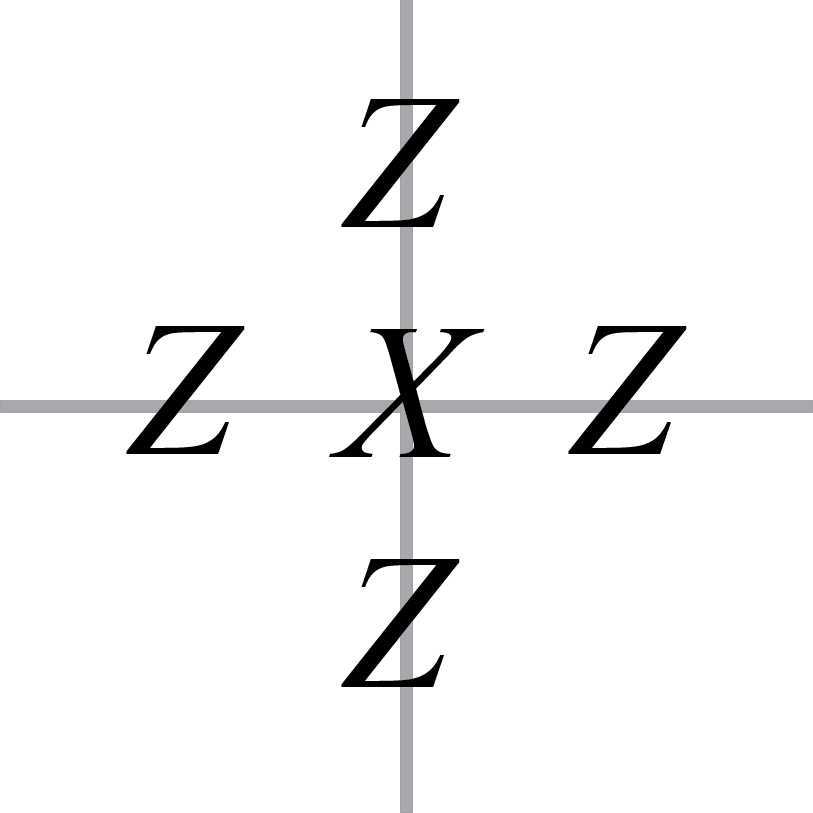}}},
\end{align}
\end{subequations}
and thus
\begin{subequations}
\begin{align}
\left(\bigotimes _{\text{all }j} C_j\right) R_i \left(\bigotimes_{\text{all }j} C_j\right) &=\  
\vcenter{\hbox{\includegraphics{IXI.png}}}, \\[2mm]
\left(\bigotimes_{\text{all }j} C_j\right) V_i \left(\bigotimes_{\text{all }j} C_j\right) &=\ 
\vcenter{\hbox{\includegraphics{ZZXZZ.png}}}, \\[2mm]
\left(\bigotimes_{\text{all }j} C_j\right) B_p \left(\bigotimes_{\text{all }j} C_j\right) &=\  
\vcenter{\hbox{\includegraphics{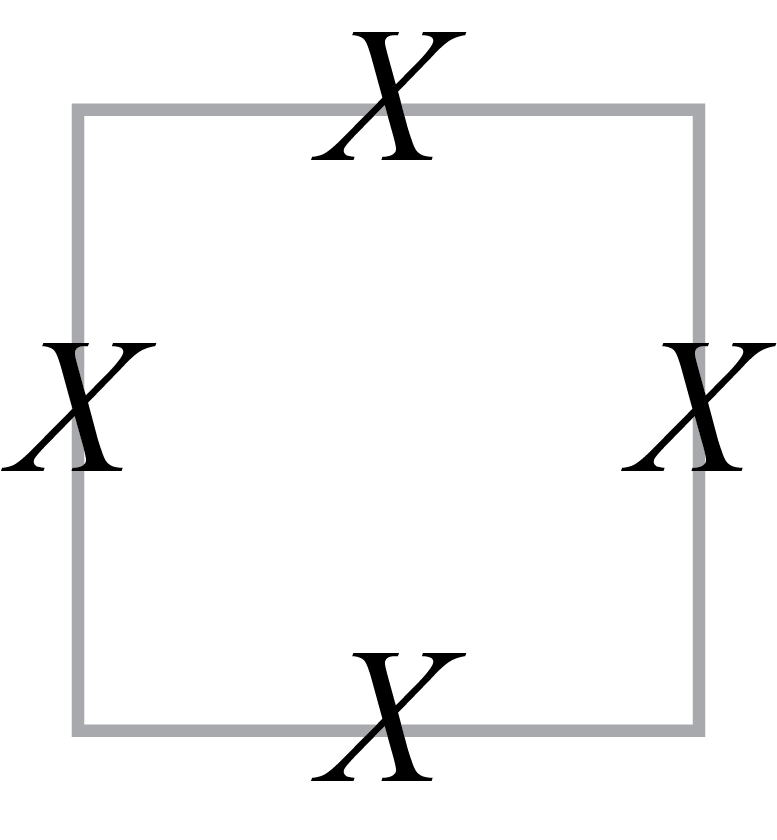}}}.
\end{align}
\end{subequations}
Using these equalities, we can easily confirm that the ground state is 
\begin{align}
|\text{GS}\rangle = \left(\bigotimes_{\text{all }j} C_j\right) \Big( |\text{TCGS}\rangle \otimes |\text{all vertices +} \rangle \Big),
\end{align}
where $|\text{TCGS}\rangle$ is the toric-code ground state for all link qubits and $|\text{all vertices +}\rangle$ is the state where all vertex qubits are in the state $|+\rangle$. Note that any such circuit necessarily violates a rotational symmetry of the lattice, so the model $H$ should be viewed as a crystalline symmetry-enriched variant of the toric code \cite{ohyama2026parameterizedfamiliestoriccode}.

\begin{figure}[t]
\includegraphics[width=\columnwidth]{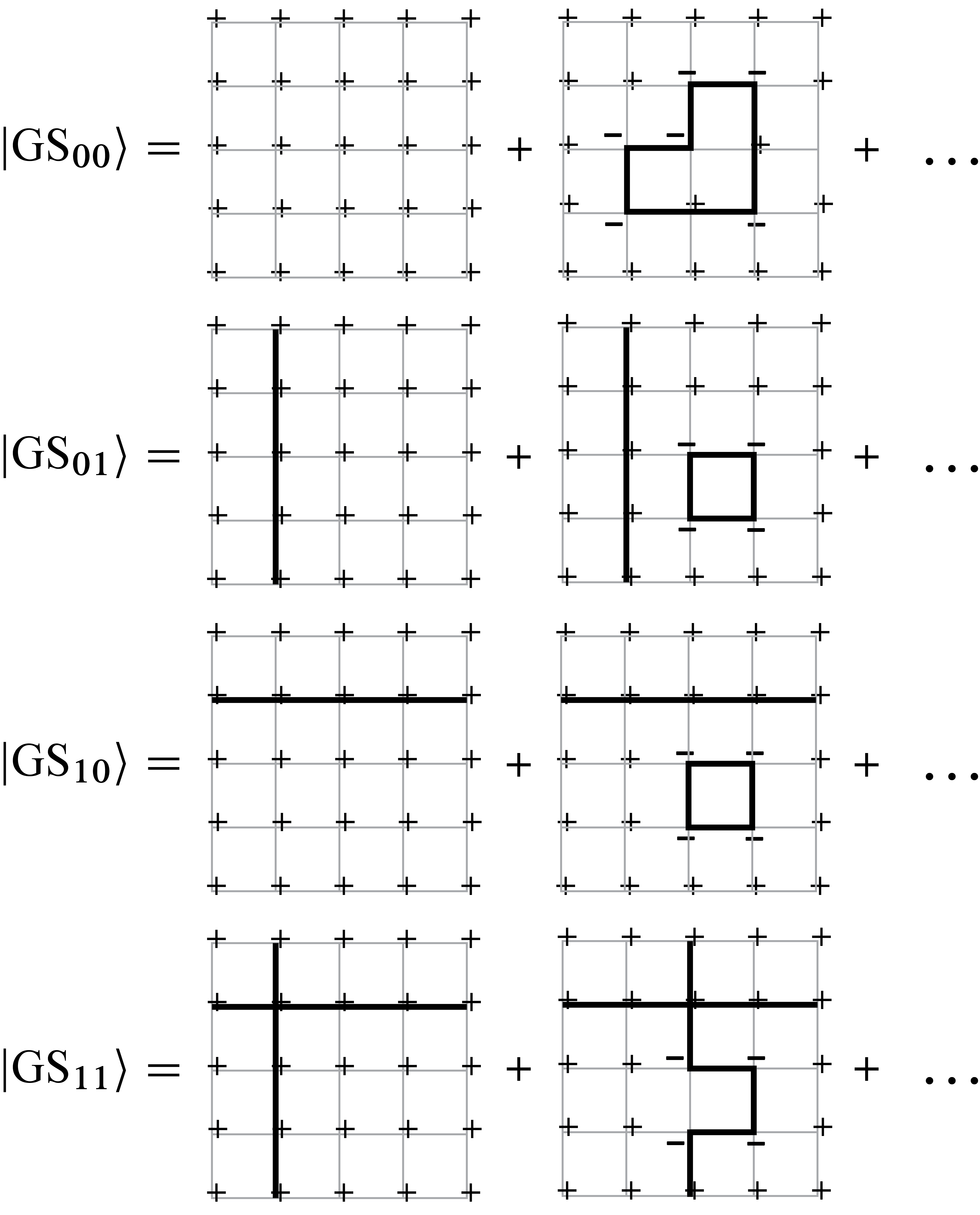}
\caption{\label{fig:GS} The ground states of the Hamiltonian \cref{eq:qec_Hamiltonian}.}
\end{figure}

To visualize the ground state, we use a bold black link for the link qubit in $\left|\downarrow\right\rangle$. As we know, the bold black links must form closed loops in $|\text{TCGS}\rangle$ and are classified into four homology classes, which correspond to four states of two logical qubits. For the vertex qubits, we simply use $+$ or $-$ to indicate their states. By \cref{eq:qec_C}, the operator $C_j$ returns $-1$ only if the vertex $j$ is in $|-\rangle$ and its left and right links are antialigned. This is a situation that only occurs at the corner of a TCGS loop. Together, the four ground states are shown in \cref{fig:GS}.

\subsection{Excited states}

Although the ground states of the model are almost identical to those of the toric code, this model has a different structure of excitations, which will affect its error correction properties.

The errors on physical qubits turn a ground state into an excited state. We say that there is an $r$ particle at vertex $i$ if $R_i|\Psi\rangle = -|\Psi\rangle$, a $v$ particle at vertex $i$ if $V_i|\Psi\rangle = -|\Psi\rangle$, and an $m$ vortex at plaquette $p$ if $B_p|\Psi\rangle = -|\Psi\rangle$.

\begin{figure*}
\includegraphics[width=0.8\textwidth]{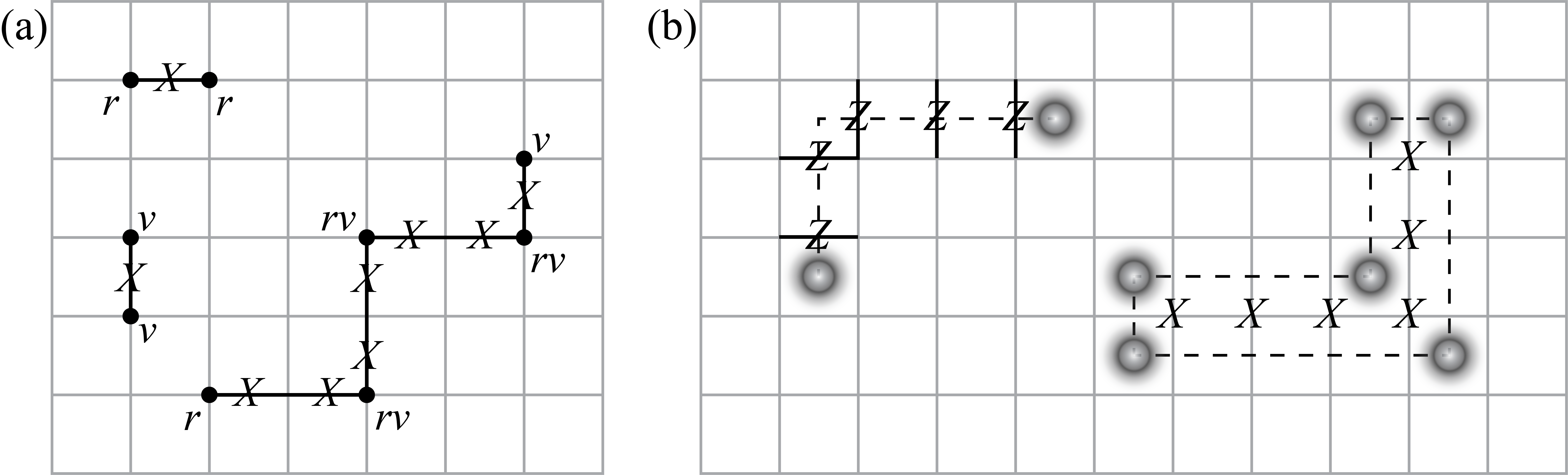}
\caption{\label{fig:creation} Creation of vertex excitations and plaquette excitations. }
\vspace*{\baselineskip}
\includegraphics[width=\textwidth]{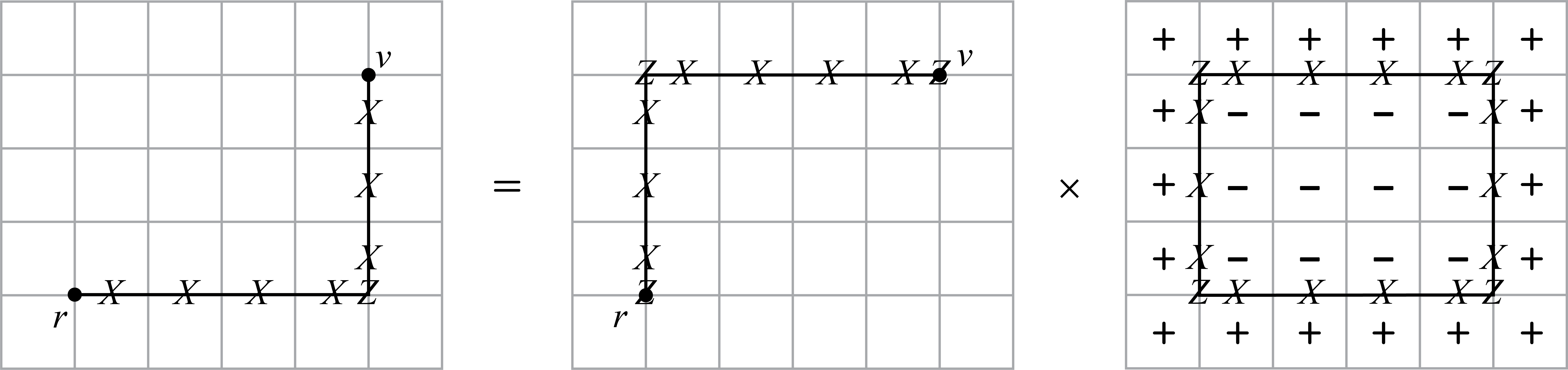}
\caption{\label{fig:rv_deformation} Examples showing the relations between two error configurations that create the same vertex syndrome. The $\pm$ signs are the corresponding SM model defined on the dual lattice.} 
\vspace*{2\baselineskip}
\includegraphics[width=0.85\textwidth]{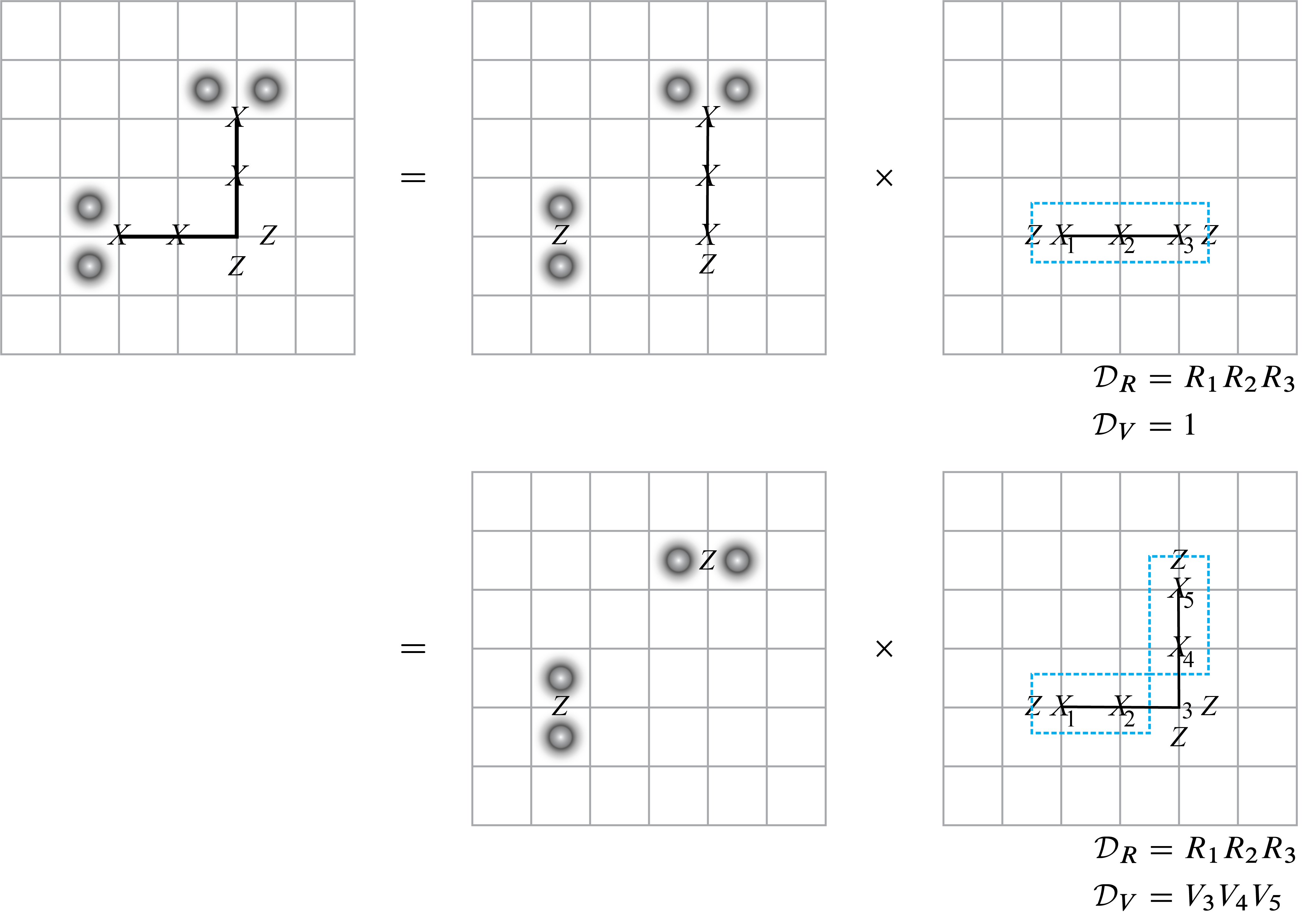}
\caption{\label{fig:m_deformation} Examples showing the relations between two error configurations that create the same plaquette syndrome. The dashed loops on the dual lattice are the results of multiplying $\mathcal{D}_R$ and $\mathcal{D}_V$.} 
\end{figure*}

The $r$ and $v$ particles are called the vertex excitations and can be created in the following two ways: 
\begin{itemize}
\item An $X$ error on a horizontal (vertical) link creates a pair of $r$ particles ($v$ particles) at the endpoints. If the $X$ errors form a string, then a pair of $r$ particles ($v$ particles) is created at every horizontal (vertical) portion of the string, and thus a composite particle consisting of an $r$ and a $v$ is created at every corner of the string. Refer to \cref{fig:creation} (a). 
\item A $Z$ error on a vertex creates a composite particle consisting of an $r$ and a $v$. 
\end{itemize}
Two error configurations, $\mathcal{E}$ and $\mathcal{E}'$, that generate the same syndrome of vertex excitations are related by
\begin{align}
\mathcal{E}' = \mathcal{E}\,\mathcal{D}_B \ \text{ with }\ \mathcal{D}_B = \bigotimes_p B_p,
\end{align}
where the tensor product is taken for any set of plaquettes. So, each connected component of $\mathcal{D}_B$ is a loop whose links have the $X$ errors and corner vertices have the $Z$ errors. See \cref{fig:rv_deformation} for an example. Any overlapping segments of two loops cancel out because $X^2 = Z^2 = 1$. 

The $m$ vortex is also called the plaquette excitation and can be created in the following two ways:
\begin{itemize}
\item A $Z$ error on a link creates a pair of $m$ vortices in the two plaquettes containing the link. If a set of $Z$ errors is crossed by a string on the dual lattice, then a pair of $m$ vortices is created in the plaquettes containing the two endpoints of the string. Refer to \cref{fig:creation} (b).
\item An $X$ error on a vertex creates four $m$ vortices, one for every plaquette containing the vertex. For a set of $X$ vertex errors, the $m$ vortices are created at the corners of the smallest dual-lattice loop enclosing these vertices. Refer to  \cref{fig:creation} (b).
\end{itemize}
Two error configurations, $\mathcal{E}$ and $\mathcal{E}'$, that generate the same syndrome of plaquette excitations are related by
\begin{align}
& \mathcal{E}' = \mathcal{E}\,\mathcal{D}_R \mathcal{D}_V \nonumber \\
& \text{with }\ \mathcal{D}_R = \bigotimes_i R_i, \quad \mathcal{D}_V= \bigotimes_j V_j,
\end{align}
where the tensor products are taken over any set of vertices. Each connected component of $\mathcal{D}_R$ ($\mathcal{D}_V$) is a loop on the dual lattice, for which every enclosed vertex has an $X$ error and every crossed horizontal (vertical) link has a $Z$ error. Again, any overlapping segments of two loops cancel out. See \cref{fig:m_deformation} for examples.

\section{Correspondence to SM models}

Now we introduce the error models and map the QEC problems to SM problems.

The basic single-qubit errors are the bit-flip error and the phase-flip error, and they are described by the Pauli $X$ and $Z$ operators, respectively. The Pauli $Y$ operator is then a combination of both as $Y = iXZ$. Let the probability of the $X$ component be $p$ and that of the $Z$ component be $q$. Assuming they are independent, the noisy channel can be written as
\begin{align}
\rho \ \mapsto\ &(1-p)(1-q)\rho \nonumber \\
&+ p(1-q) X\rho X + (1-p)q Z\rho Z + pq\, Y\rho Y, \label{eq:noisy_channel}
\end{align}
where $\rho$ is the (reduced) density operator of a physical qubit. This states that the probability of a physical qubit suffering a bit-flip error is $p(1-q)$, while the probability of a phase-flip error is $(1-p)q$. We also assume that the errors on different qubits are independent. This is the noisy channel we will consider. Together with the fact that our code is locally Clifford-equivalent to a CSS code and that each of $\mathcal{D}_R$, $\mathcal{D}_V$, and $\mathcal{D}_B$ only deforms its respective syndrome type, the decoding problem can be split into two independent problems, one for the vertex syndromes and the other for the plaquette syndromes. For simplicity, we also assume a perfect measurement of the syndrome so that we do not need to consider successive measurements in time.

Given a measured syndrome, which is caused by an unknown error configuration $\mathcal{E}$, the decoder applies $\mathcal{E}' = \mathcal{E} \mathcal{D}$ to remove the syndrome. The QEC is successful if $\mathcal{D}$ is a product of stabilizers. The QEC may fail, however, if $\mathcal{D}$ realizes a logical operator wrapping a cycle of the torus. The probability of a failed QEC is therefore
\begin{align}
P(\text{failure}) = \sum_{\mathcal{E}} P(\mathcal{E}) \frac{\displaystyle\sum_{\text{nontrivial }\mathcal{D}} \hspace*{-4mm} P(\mathcal{E}\mathcal{D} \mid \mathcal{E})}{\displaystyle\sum_{\text{all }\mathcal{D}} P(\mathcal{E}\mathcal{D} \mid \mathcal{E})}. \label{eq:failure_prob}
\end{align}
We want to find the range of $p$ (the probability of $X$) and $q$ (the probability of $Z$) so that the probability of a failed QEC is zero in the limit of infinite system size. 

\subsection{Vertex syndromes}

We first consider the syndromes consisting only of vertex excitations. We will show that the decoding problem corresponds to a random-bond Ising model with random plaquette interactions \cref{eq:SM_Hamiltonian} with parameters given in \cref{eq:Nishimori}.

An error configuration of vertex syndromes is written as $\mathcal{E} = \mathcal{X}\otimes\mathcal{Z}$, where $\mathcal{X}$ is the loop configuration of $X$ errors on links and $\mathcal{Z}$ is the configuration of $Z$ errors at the corner vertices of these loops. Let $n_\mathcal{X}(\ell) = 1$ if $\ell\in\mathcal{X}$, and let $n_\mathcal{X}(\ell)=0$ if not. Let $n_\mathcal{Z}(i) = 1$ if $i\in\mathcal{Z}$, and let $n_\mathcal{Z}(i)=0$ if not. Then the probabilities of whether the link $\ell$ and the vertex $i$ are erroneous can be written as 
\begin{subequations}
\begin{align}
P(\ell\in \mathcal{X}) = p^{n_\mathcal{X}(\ell)} (1-p)^{1-n_\mathcal{X}(\ell)} \propto \left( \frac{p}{1-p} \right)^{n_\mathcal{X}(\ell)},\\
P(i\in \mathcal{Z}) = q^{n_\mathcal{Z}(\ell)} (1-q)^{1-n_\mathcal{Z}(\ell)} \propto \left( \frac{q}{1-q} \right)^{n_\mathcal{Z}(i)},
\end{align}
\end{subequations}
where $p$ and $q$ are the parameters of the error model \cref{eq:noisy_channel}, and the probability of an error configuration is 
\begin{align}
P(\mathcal{E}) \propto\ 
\prod_\ell \left( \frac{p}{1-p} \right)^{n_\mathcal{X}(\ell)} \prod_i \left( \frac{q}{1-q} \right)^{n_\mathcal{Z}(i)}. \label{eq:prob_error_rv}
\end{align}
Another error configuration $\mathcal{E}' =\mathcal{X}' \otimes \mathcal{Z}'$ that causes the same syndrome is related by 
\begin{subequations}
\begin{align}
\mathcal{X}' = \mathcal{X} \overline{\mathcal{D}}, \\
\mathcal{Z}' = \mathcal{Z} \widehat{\mathcal{D}},
\end{align}
\end{subequations}
where $\overline{\mathcal{D}}$ denotes the links and $\widehat{\mathcal{D}}$ denotes the corners of $\mathcal{D}$, and $\mathcal{D} = \overline{\mathcal{D}}\otimes \widehat{\mathcal{D}} = \bigotimes_p B_p$. A link $\ell$ belongs to $\mathcal{X}'$ if it belongs to $\mathcal{X}$ or $\overline{\mathcal{D}}$ but not both; a vertex $i$ belongs to $\mathcal{Z}'$ if it belongs to $\mathcal{Z}$ or $\widehat{\mathcal{D}}$ but not both. So,
\newpage\vspace*{-2\baselineskip}
\begin{subequations}
\begin{align}
n_{\mathcal{X}'} (\ell) &= \delta_{n_\mathcal{X}(\ell) + n_{\overline{\mathcal{D}}}(\ell), 1} \nonumber\\
&= n_\mathcal{X}(\ell) + n_{\overline{\mathcal{D}}}(\ell) - 2 n_\mathcal{X}(\ell) n_{\overline{\mathcal{D}}}(\ell), \\
n_{\mathcal{Z}'} (i) &= \delta_{n_\mathcal{Z}(i) + n_{\widehat{\mathcal{D}}}(i), 1} \nonumber\\
&= n_\mathcal{Z}(i) + n_{\widehat{\mathcal{D}}}(i) - 2n_\mathcal{Z}(i) n_{\widehat{\mathcal{D}}}(i) ,
\end{align}
\end{subequations}
where $\delta$ is the Kronecker delta. Then we have the following conditional probabilities:
\begin{subequations} 
\begin{align}
P(\ell \in \mathcal{X}' \mid \ell\in \mathcal{X}) 
& \propto \left( \frac{p}{1-p} \right)^{n_{\mathcal{X}'}(\ell)}\bigg|_{n_{\mathcal{X}}(\ell)=1} \nonumber \\
&= \left(\frac{p}{1-p}\right)^{1-n_{\overline{\mathcal{D}}}(\ell)} 
\propto \left(\frac{1-p}{p}\right)^{n_{\overline{\mathcal{D}}}(\ell)}, \\
P(\ell \in \mathcal{X}' \mid \ell\notin \mathcal{X})
& \propto \left( \frac{p}{1-p} \right)^{n_{\mathcal{X}'}(\ell)}\bigg|_{n_{\mathcal{X}}(\ell)=0} \nonumber\\
&= \left(\frac{p}{1-p}\right)^{n_{\overline{\mathcal{D}}}(\ell)}, \\
P(i \in \mathcal{Z}' \mid i\in \mathcal{Z}) 
& \propto \left( \frac{q}{1-q} \right)^{n_{\mathcal{Z}'}(i)}\bigg|_{n_{\mathcal{Z}}(i)=1}\nonumber\\
&= \left(\frac{q}{1-q}\right)^{1-n_{\widehat{\mathcal{D}}}(i)} 
\propto \left(\frac{1-q}{q}\right)^{n_{\widehat{\mathcal{D}}}(i)}, \\
P(i \in \mathcal{Z}' \mid i\notin \mathcal{Z}) 
& \propto \left( \frac{q}{1-q} \right)^{n_{\mathcal{Z}'}(i)}\bigg|_{n_{\mathcal{Z}}(i)=0}\nonumber\\
&= \left(\frac{q}{1-q}\right)^{n_{\widehat{\mathcal{D}}}(i)}.
\end{align}
\end{subequations}
Together, 
\begin{align*}
P(\mathcal{E}'\mid \mathcal{E}) 
= &\prod_{\ell\in \mathcal{X}} P(\ell \in \mathcal{X}'\mid \ell\in \mathcal{X}) \ \prod_{\ell\notin \mathcal{X}} P(\ell \in \mathcal{X}'\mid \ell\notin \mathcal{X}) \\
&\prod_{i\in \mathcal{Z}} P(i \in \mathcal{Z}'\mid i\in \mathcal{Z}) \ \prod_{i\notin \mathcal{Z}} P(i \in \mathcal{Z}'\mid i\notin \mathcal{Z}) \\
\propto &\prod_{\ell\in \mathcal{X}} \left(\frac{1-p}{p}\right)^{n_{\overline{\mathcal{D}}}(\ell)} \ 
\prod_{\ell\notin \mathcal{X}} \left(\frac{p}{1-p}\right)^{n_{\overline{\mathcal{D}}}(\ell)}\\
&\prod_{i\in \mathcal{Z}} \left(\frac{1-q}{q}\right)^{n_{\widehat{\mathcal{D}}}(i)} \ 
\prod_{i\notin \mathcal{Z}} \left(\frac{q}{1-q}\right)^{n_{\widehat{\mathcal{D}}}(i)} \\
= &\prod_{\ell\in \mathcal{X}} \left(\frac{1-p}{p}\right)^{n_{\overline{\mathcal{D}}}(\ell)-\frac{1}{2}} \ 
\prod_{\ell\notin \mathcal{X}} \left(\frac{p}{1-p}\right)^{n_{\overline{\mathcal{D}}}(\ell)-\frac{1}{2}}\\
& \prod_{i\in \mathcal{Z}} \left(\frac{1-q}{q}\right)^{n_{\widehat{\mathcal{D}}}(i)-\frac{1}{2}} \ 
\prod_{i\notin \mathcal{Z}} \left(\frac{q}{1-q}\right)^{n_{\widehat{\mathcal{D}}}(i)-\frac{1}{2}} ,
\end{align*}
where a factor with power $1/2$ was canceled. Make the substitution
\begin{subequations}\label{eq:error_rv_d}
\begin{align}
d(\ell) &= -2n_{\overline{\mathcal{D}}}(\ell) + 1, \\
d(i) &= -2n_{\widehat{\mathcal{D}}}(i) + 1 ,
\end{align}
\end{subequations}
so that $\{ \ell: d(\ell) = -1 \}$ is the loop configuration $\overline{\mathcal{D}}$ and $\{ i: d(i) = -1 \}$ is the corner configuration $\widehat{\mathcal{D}}$. Also, make the substitution
\begin{subequations}\label{eq:error_rv_eta}
\begin{align}
\eta(\ell) &= -2n_{\mathcal{X}}(\ell) + 1, \\
\eta(i) &= -2n_{\mathcal{Z}}(i) + 1 ,
\end{align}
\end{subequations}
so that the $\eta$ configuration is the $\mathcal{E}$ configuration: $\{ \ell: \eta(\ell) = -1 \}$ and $\{ i: \eta(i) = -1 \}$ are the erroneous links and vertices, respectively. Now we rewrite the probability in the form of the thermal probability:
\begin{align}
e^{-2\beta} = \frac{p}{1-p}, \qquad 
e^{-2\beta K} = \frac{q}{1-q}. \label{eq:Nishimori}
\end{align}
The first equation in \cref{eq:Nishimori} is called the Nishimori surface in the three-dimensional space of $p$, $q$, and $\beta$ (in the case of the toric code it is called the Nishimori line as it only has $p$ and $\beta$ \cite{Dennis2002}), while the second equation determines the value of $K$. Putting Eqs. \eqref{eq:error_rv_d}--\eqref{eq:Nishimori} together, the conditional probability
\begin{align}
P(\mathcal{E}'\mid \mathcal{E}) \propto e^{-\beta H_{\eta}} \label{eq:conditional_prob}
\end{align}
with the Hamiltonian 
\begin{align}
H_{\eta} = -\sum_\ell \eta(\ell) d(\ell) - K \sum_i \eta(i) d(i)
\end{align}
and the restriction that $\{ \ell: d(\ell) = -1 \}$ must form loops and $\{ i: d(i) = -1 \}$ must be the corners of these loops.
Because of this restriction, the loops of $d(\ell)=-1$ can be understood as the domain walls (DWs) of an Ising model defined on the dual-lattice vertices \footnote{Note that, unlike those in \cite{Dennis2002}, the DWs we defined occur where neighboring Ising spins are antialigned. So, they always form closed loops and may or may not raise the energy depending on the value of $\eta(\ell)$. The DWs defined in \cite{Dennis2002} are the superposition of $d(\ell)$ and $\eta(\ell)$ and thus may not be closed but always raise the energy.}. Moreover, our model concerns $d(i)$, the corners of DWs, where one or three surrounding Ising spins are negative. Refer to \cref{fig:rv_deformation}. Thus, we have the following transformation:
\begin{subequations}
\begin{align}
d(\ell) &= \sigma_a \sigma_b,\\
d(i) &= \sigma_a \sigma_b \sigma_c \sigma_d,
\end{align}
\end{subequations}
where $a$ and $b$ are the nearest neighbors separated by $\ell$, and $a$, $b$, $c$, and $d$ are the centers of the plaquettes containing the vertex $i$. The Hamiltonian is rewritten as
\begin{align}
H_{\eta} = -\sum_{\text{nn}(a,b)} \eta_{ab}\, \sigma_a \sigma_b  - K \sum_{\text{plaq}(a,b,c,d)} \eta_{abcd}\, \sigma_a \sigma_b \sigma_c \sigma_d ,\label{eq:SM_Hamiltonian}
\end{align}
where ``nn'' stands for the nearest neighbors. This is the random-bond Ising model with random-plaquette four-spin interaction. An error configuration $\mathcal{E}$ corresponds to a realization of disordered bonds and plaquettes, i.e., an $\eta$ configuration, while a deformation $\mathcal{D}$ corresponds to a DW configuration. 

Plugging \cref{eq:conditional_prob} back into the failure probability \cref{eq:failure_prob} and identifying $\mathcal{E}$ with $\eta$, we get
\newpage\vspace*{-2.5\baselineskip}
\begin{align}
P(\text{failure}) &= \sum_{\eta} P( \eta ) \frac{\displaystyle\sum_{\text{nontrivial }\mathcal{D}} \hspace*{-4mm} e^{-\beta H_\eta} }{\displaystyle\sum_{\text{all }\mathcal{D}}e^{-\beta H_\eta} } \nonumber\\
&= \sum_{\eta} P(\eta) \frac{\displaystyle\sum_{h\neq 0} Z_{\eta,h} }{Z_{\eta, 0} + \displaystyle\sum_{h\neq 0} Z_{\eta,h} } \nonumber \\ 
&= \sum_{\eta} P(\eta) \frac{\displaystyle\sum_{h\neq 0} e^{-\beta (F_{\eta,h} - F_{\eta,0})} }{1 + \displaystyle\sum_{h\neq 0} e^{-\beta (F_{\eta,h} - F_{\eta,0})} } ,
\end{align}
where $h$ is the homology class of $\mathcal{D}$ and $h=0$ is the trivial class; $Z_{\eta, h}$ and $F_{\eta,h}$ are the partition function and the free energy of the homology class $h$. Define $\Delta F_\eta$ as the minimum free-energy difference between the nontrivial and the trivial DW classes,
\begin{align}
\Delta F_\eta = \min_{h\neq 0} (F_{\eta,h} - F_{\eta, 0}) ,
\end{align} 
then we have
\begin{align}
P(\text{failure}) 
& \leqslant \sum_{\eta} P(\eta) \sum_{h\neq 0} e^{-\beta (F_{\eta,h} - F_{\eta,0})} \nonumber\\
& \leqslant 3 \sum_{\eta} P(\eta) e^{-\beta \Delta F_\eta} . \label{eq:fail_to_ferro}
\end{align}
In the case $p=q=0$, i.e., a disorder-free system, $\Delta F$ of the ferromagnetic phase should diverge in the thermodynamic limit, $L\rightarrow\infty$, because $\Delta F$ is dominated by the internal energy $\langle \Delta E\rangle$ of a nontrivial DW, which is $\langle \Delta E\rangle = 2L + 2K\langle C\rangle$ with 
\begin{align}
\langle C\rangle &= \frac{\displaystyle\sum_{C=4, 6, ..., L} C e^{-\beta 2KC} }{\displaystyle\sum_C e^{-\beta 2KC}} \nonumber\\
&= 4 + \frac{2}{e^{4\beta K}-1} - \frac{L-2}{e^{2\beta K(L-2)} - 1}
\end{align}
and thus diverges as $L\rightarrow\infty$. For the general case, $\Delta F_\eta$, however, depends on the specific disorder realization $\eta$. Hence, the same ferromagnetic phase must be identified as the regime where the probability of the disorder realizations $\eta$ for which $\Delta F_\eta \rightarrow \infty$ as $L\rightarrow\infty$ is close to 100\%. Then, by \cref{eq:fail_to_ferro}, the probability of failure is zero in the ferromagnetic phase. Equivalently, this ferromagnetic phase can also be characterized by a nonzero magnetization squared \cite{Dennis2002},
\begin{align}
\mathbb{E}(\langle M^2\rangle) = \frac{1}{N^2} \sum_{i,j} \mathbb{E} (\langle s_i s_j\rangle), \label{eq:magnet2}
\end{align}
which then vanishes in the paramagnetic phase.

\subsection{Plaquette syndromes}

We now consider the syndromes consisting only of plaquette excitations. We will show that the decoding problem corresponds to two layers of perpendicular one-dimensional Ising models with interlayer coupling \cref{eq:SM_Hamiltonian_dual} with the same parameters given in \cref{eq:Nishimori}.

An error configuration of plaquette syndromes is written as $\mathcal{E} = \mathcal{X}\otimes\mathcal{Z}$, where $\mathcal{X}$ is the configuration of $X$ errors on vertices and $\mathcal{Z}$ is the configuration of $Z$ errors on links. Another error configuration $\mathcal{E}' =\mathcal{X}' \otimes \mathcal{Z}'$ that causes the same syndrome is related by 
\begin{subequations}
\begin{align}
\mathcal{X}' = \mathcal{X}\, \widehat{\mathcal{D}}_R \widehat{\mathcal{D}}_V, \\
\mathcal{Z}' = \mathcal{Z}\, \overline{\mathcal{D}}_R \overline{\mathcal{D}}_V,
\end{align}
\end{subequations}
where $\widehat{\mathcal{D}}_R$ denotes the vertices and $\overline{\mathcal{D}}_R$ denotes the horizontal links, and $\overline{\mathcal{D}}_R\otimes \widehat{\mathcal{D}}_R = \prod_i R_i$. The same applies to $V_i$ and the vertical links. A vertex $i$ belongs to $\mathcal{X}'$ if it belongs to one or three of $\mathcal{X}$, $\widehat{\mathcal{D}}_R$, and $\widehat{\mathcal{D}}_V$. A link, however, belongs to $\mathcal{Z}'$ if it belongs to only one of $\mathcal{Z}$, $\overline{\mathcal{D}}_R$, and $\overline{\mathcal{D}}_V$ since a link cannot be both horizontal and vertical; that is, $n_{\overline{\mathcal{D}}_R} (\ell) n_{\overline{\mathcal{D}}_V} (\ell) = 0$. So, 
\begin{subequations}
\begin{align}
& n_{\mathcal{X}'} (i) \nonumber\\ 
=\,& \delta_{n_\mathcal{X}(i) + n_{\widehat{\mathcal{D}}_R}(i) + n_{\widehat{\mathcal{D}}_V}(i), 1} + \delta_{n_\mathcal{X}(i) + n_{\widehat{\mathcal{D}}_R}(i) + n_{\widehat{\mathcal{D}}_V}(i), 3} \nonumber \\
=\,& n_\mathcal{X}(i) + n_{\widehat{\mathcal{D}}_R}(i) + n_{\widehat{\mathcal{D}}_V}(i) \nonumber\\
&- 2\left( n_\mathcal{X}(i) n_{\widehat{\mathcal{D}}_R}(i) + n_\mathcal{X}(i) n_{\widehat{\mathcal{D}}_V}(i) + n_{\widehat{\mathcal{D}}_R}(i) n_{\widehat{\mathcal{D}}_V}(i)\right) \nonumber\\
&+ 4n_\mathcal{X}(i) n_{\widehat{\mathcal{D}}_R}(i) n_{\widehat{\mathcal{D}}_V}(i), \\
& n_{\mathcal{Z}'} (\ell) \nonumber\\ 
=\,& n_\mathcal{Z}(\ell) + n_{\overline{\mathcal{D}}_R}(\ell) + n_{\overline{\mathcal{D}}_V}(\ell) \nonumber\\
&- 2\left( n_\mathcal{Z}(\ell) n_{\overline{\mathcal{D}}_R}(\ell) + n_\mathcal{Z}(\ell) n_{\overline{\mathcal{D}}_V}(\ell)\right).
\end{align}
\end{subequations}
Then we have the following conditional probabilities:
\begin{subequations} 
\begin{align}
P(i \in \mathcal{X}' &\mid i\in \mathcal{X}) \nonumber\\
&\propto \left( \frac{p}{1-p} \right)^{n_{\mathcal{X}'}(i)}\bigg|_{n_{\mathcal{X}}(i)=1} \nonumber\\
&= \left(\frac{p}{1-p}\right)^{1-n_{\widehat{\mathcal{D}}_R}(i)-n_{\widehat{\mathcal{D}}_V}(i)+2n_{\widehat{\mathcal{D}}_R}(i)n_{\widehat{\mathcal{D}}_V}(i) } \nonumber\\
&\propto \left(\frac{1-p}{p}\right)^{\left| n_{\widehat{\mathcal{D}}_R}(i) - n_{\widehat{\mathcal{D}}_V}(i)\right|}, \\
P(i \in \mathcal{X}' &\mid i\notin \mathcal{X}) \nonumber\\
&\propto \left( \frac{p}{1-p} \right)^{n_{\mathcal{X}'}(i)}\bigg|_{n_{\mathcal{X}}(i)=0} \nonumber\\
&= \left(\frac{p}{1-p}\right)^{\left| n_{\widehat{\mathcal{D}}_R}(i) - n_{\widehat{\mathcal{D}}_V}(i)\right|} ,\\
P(\ell \in \mathcal{Z}' & \mid \ell\in \mathcal{Z}) \nonumber \\
&\propto \left( \frac{q}{1-q} \right)^{n_{\mathcal{Z}'}(\ell)}\bigg|_{n_{\mathcal{Z}}(\ell)=1}\nonumber\\
&= \left(\frac{q}{1-q}\right)^{1-n_{\overline{\mathcal{D}}_R}(\ell)-n_{\overline{\mathcal{D}}_V}(\ell)} \nonumber\\
&\propto \left(\frac{1-q}{q}\right)^{n_{\overline{\mathcal{D}}_R}(\ell)+n_{\overline{\mathcal{D}}_V}(\ell)}, \\
P(\ell \in \mathcal{Z}' & \mid \ell\notin \mathcal{Z}) \nonumber\\
&\propto \left( \frac{q}{1-q} \right)^{n_{\mathcal{Z}'}(\ell)}\bigg|_{n_{\mathcal{Z}}(\ell)=0}\nonumber\\
&= \left(\frac{q}{1-q}\right)^{n_{\overline{\mathcal{D}}_R}(\ell)+n_{\overline{\mathcal{D}}_V}(\ell)}.
\end{align}
\end{subequations}
Together, 
\begin{align}
P(\mathcal{E}'\mid \mathcal{E}) 
=& \prod_{i\in \mathcal{X}} P(i \in \mathcal{X}'\mid i\in \mathcal{X}) \ \prod_{i\notin \mathcal{X}} P(i \in \mathcal{X}'\mid i\notin \mathcal{X}) \nonumber\\
&\prod_{\ell\in \mathcal{Z}} P(\ell \in \mathcal{Z}'\mid \ell\in \mathcal{Z}) \ \prod_{\ell\notin \mathcal{Z}} P(\ell \in \mathcal{Z}'\mid \ell\notin \mathcal{Z})\nonumber \\
\propto& \prod_{i\in \mathcal{X}} \left(\frac{1-p}{p}\right)^{\left| n_{\widehat{\mathcal{D}}_R}(i) - n_{\widehat{\mathcal{D}}_V}(i)\right| - \frac{1}{2}} \nonumber \\
&\prod_{i\notin \mathcal{X}} \left(\frac{p}{1-p}\right)^{\left| n_{\widehat{\mathcal{D}}_R}(i) - n_{\widehat{\mathcal{D}}_V}(i)\right| - \frac{1}{2}} \nonumber \\
&\prod_{\ell\in \mathcal{Z}} \left(\frac{1-q}{q}\right)^{n_{\overline{\mathcal{D}}_R}(\ell)+n_{\overline{\mathcal{D}}_V}(\ell) - 1} \nonumber \\
&\prod_{\ell\notin \mathcal{Z}} \left(\frac{q}{1-q}\right)^{n_{\overline{\mathcal{D}}_R}(\ell)+n_{\overline{\mathcal{D}}_V}(\ell) - 1},
\end{align}
where a factor with power $1/2$ and a factor with power 1 were canceled. Make the substitution
\begin{subequations}\label{eq:error_m_d}
\begin{align}
d(i) &= -2 \left| n_{\widehat{\mathcal{D}}_R}(i) - n_{\widehat{\mathcal{D}}_V}(i)\right| + 1, \label{eq:error_m_d_vertex}\\
d(\ell) &= -2n_{\overline{\mathcal{D}}_R}(\ell) + 1 \ \text{ for horizontal links},\\
d(\ell) &= -2n_{\overline{\mathcal{D}}_V}(\ell) + 1 \ \text{ for vertical links}
\end{align}
\end{subequations}
so that $\{ \ell: d(\ell) = -1 \}$ is $\overline{\mathcal{D}}_R \otimes \overline{\mathcal{D}}_V$ and $\{ i: d(i) = -1 \}$ is $\widehat{\mathcal{D}}_R \widehat{\mathcal{D}}_V$, i.e., the resulting configuration of $X$ on vertices. Also, make the substitution
\begin{subequations} \label{eq:error_m_eta}
\begin{align}
\eta(i) &= -2n_{\mathcal{X}}(i) + 1,\\
\eta(\ell) &= -2n_{\mathcal{Z}}(\ell) + 1
\end{align}
\end{subequations}
so that the $\eta$ configuration is the $\mathcal{E}$ configuration: $\{ i: \eta(i) = -1 \}$ and $\{ \ell: \eta(\ell) = -1 \}$ are the erroneous vertices and links, respectively. By the same definition of the Nishimori surface \cref{eq:Nishimori}, the conditional probability becomes $P(\mathcal{E}'\mid \mathcal{E}) \propto e^{-\beta H_\eta}$ with the Hamiltonian 
\begin{align}
H_{\eta} = &-\sum_i \eta(i) d(i) \nonumber \\
&- K \hspace*{-3mm}\sum_{\text{horizontal }\ell}\hspace*{-4mm} \eta(\ell) d(\ell) - K \hspace*{-2mm} \sum_{\text{vertical }\ell}\hspace*{-2mm} \eta(\ell) d(\ell) 
\end{align}
and the restriction that $\{ \ell: d(\ell) = -1 \} \cup \{ i: d(i) = -1 \}$ must be $\bigotimes_i R_i \bigotimes_j V_j$. The absolute value in \cref{eq:error_m_d_vertex} merely says that $d(i)=+1$ if both $R_i$ and $V_i$ act there, and $d(i)=-1$ if one of $R_i$ and $V_i$ acts there. This implies that we can define \textit{two species} of Ising spins at a vertex on the direct lattice, $\sigma_i$ and $\tau_i$, so that $u(i)=-1$ corresponds to two opposite on-site spins. A horizontal (vertical) link with $d(\ell)=-1$ connects a horizontal pair of antialigned $\sigma$'s (a vertical pair of antialigned $\tau$'s). Thus, we have the following transformation:
\begin{subequations}
\begin{align}
d(i) &= \sigma_{x,y} \tau_{x,y} ,\\
d(\ell) &= \sigma_{x,y} \sigma_{x+1,y} \ \text{ for horizontal $\ell$} ,\\
d(\ell) &= \tau_{x,y} \tau_{x, y+1} \ \text{ for vertical $\ell$} ,
\end{align}
\end{subequations}
where $(x,y)$ are the Cartesian coordinates of the vertex $i$. Then the Hamiltonian is 
\begin{align}
H_{\eta} = &-\sum_{x,y} \eta_{x,y} \sigma_{x,y} \tau_{x,y} \nonumber \\
&- K \sum_{x,y} \eta_{(x,y)(x,y+1)} \tau_{x,y} \tau_{x, y+1} \nonumber \\
&- K\sum_{x,y} \eta_{(x,y)(x+1,y)} \sigma_{x,y} \sigma_{x+1, y}. \label{eq:SM_Hamiltonian_dual}
\end{align}
This is a two-layer Ising model: $\eta_{x,y}\sigma_{x,y}$ is the interlayer coupling; the $\tau$-layer is a collection of independent one-dimensional Ising models in the $y$-direction, and the $\sigma$-layer is a collection of independent one-dimensional Ising models in the $x$-direction. This model is different from the model for decoding vertex syndromes, that is \cref{eq:SM_Hamiltonian}, but the two are dual to each other in case of no errors, i.e., $p=q=0$. See Appendix \ref{app:duality}.

\section{Numerical analysis of the error thresholds}
As we showed in Section II, the range of physical error rates, $p$ and $q$, for a successful QEC is where the Nishimori surface is in the ferromagnetic phase. This means that the numerical analysis only needs to scan the Nishimori surface. However, here we present a more complete study of the phase diagram. 

The SM model, \cref{eq:SM_Hamiltonian}, has never been studied so far except for the disorder-free version which was numerically tested for the self-learning Monte Carlo method \cite{Fu2017}. To handle the disorder, here we adopt the parallel tempering Markov-chain Monte Carlo (MCMC) algorithm \cite{Hukushima1996}: for each disorder realization, we initialize a list of different spin configurations at different temperatures; then we exchange, according to detailed balance, the two at adjacent temperatures after performing a certain number of Metropolis updates. To avoid the possibility that a configuration is moved directly to the highest temperature in one implementation of parallel tempering, we propose the exchange of them at $T_n$ and $T_{n+1}$ only for even $n$ in one implementation and only for odd $n$ in the next implementation, and repeat in this even-odd alternating way. In addition, the temperature ladder is crucial for the performance of parallel tempering. A good temperature ladder gives a relatively uniform exchange rate between each pair of adjacent temperatures and ensures that every configuration can complete a full round trip. We tested the adaptive ladder generation algorithm given by \cite{Katzgraber2006} and found that it can produce a distant separation between the highest temperature and the others, which leads to poor dynamics in the temperature space. We also tested the uniform $\beta$, the geometric $T$, and the geometric $\beta$ ladders, but they are worse than the uniform $T$ ladder. Therefore, we use the uniform $T$ ladder in our work, yet we cannot conclude that it is the best for this SM model.

\begin{figure}
\includegraphics[width=\columnwidth]{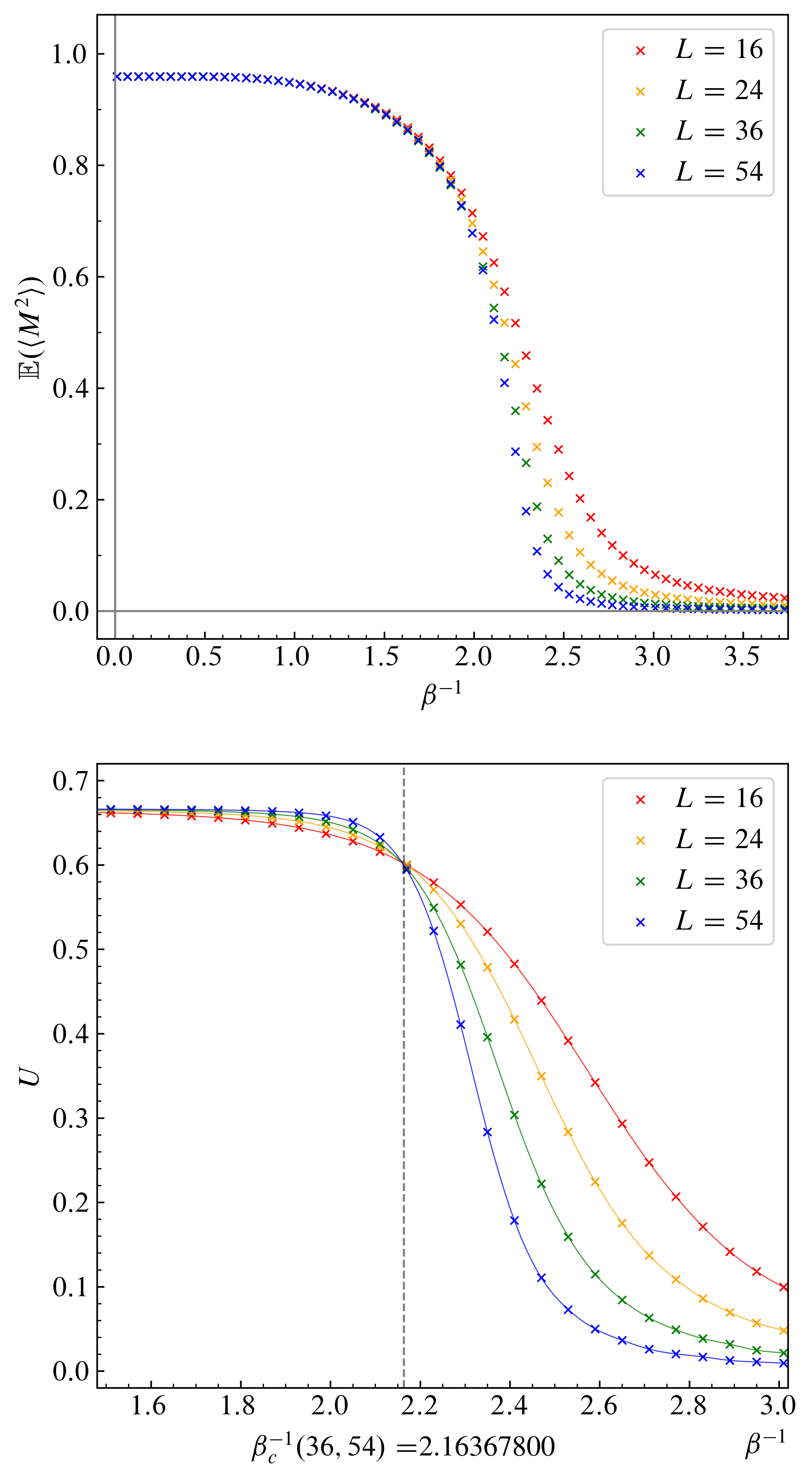}
\caption{\label{fig:plot_m} Magnetization squared $\mathbb{E}(\langle M^2\rangle)$ and the Binder cumulant $U$ at $p=q=0.1$ of the SM model \cref{eq:SM_Hamiltonian}.} 
\end{figure}

The critical temperature $T_c$ is found by considering the crossing point of the Binder cumulants of two system sizes, $L$ and $r L$. The magnetization (per site) is defined as 
\begin{align}
M = \frac{1}{N}\sum_i \sigma_i\, ,
\end{align}
and the disorder-averaged Binder cumulant is defined as \cite{Privman1990}
\begin{align}
U(T) = 1 - \frac{\mathbb{E}(\langle M^4\rangle)}{3\,\mathbb{E}(\langle M^2\rangle)^2} ,
\end{align}
where $\langle\cdot\rangle$ represents the thermal average and $\mathbb{E}(\cdot)$ represents the disorder average. Near $T_c$, it scales as
\begin{align}
U_L(T) = U_0 + a(T-T_c) L^{1/\nu} + bL^{-\omega} + \cdots. \label{eq:binder_scaling}
\end{align}
At the crossing point, $U_L(T_\text{cross}) = U_{rL}(T_\text{cross})$, \cref{eq:binder_scaling} gives
\begin{align}
T_\text{cross} (L, rL) = T_c + a L^{-\lambda}.
\end{align}
This shows that $T_c$ is $T_\text{cross}$ in the thermodynamic limit, $L\rightarrow\infty$. We obtain the estimated $T_c$ by a weighted linear fit of $T_\text{cross}$ of $(L,1.5L) = (16,24)$, $(24,36)$, and $(36,54)$ against $L^{-1}$, and we obtain the error bar representing the 95\% confidence interval by bootstrap resampling 1080 disorder realizations. A sample result of $\mathbb{E}(\langle M\rangle^2)$ and $U(T)$ is given by \cref{fig:plot_m}. 

\begin{figure}[h]
\includegraphics[width=\columnwidth]{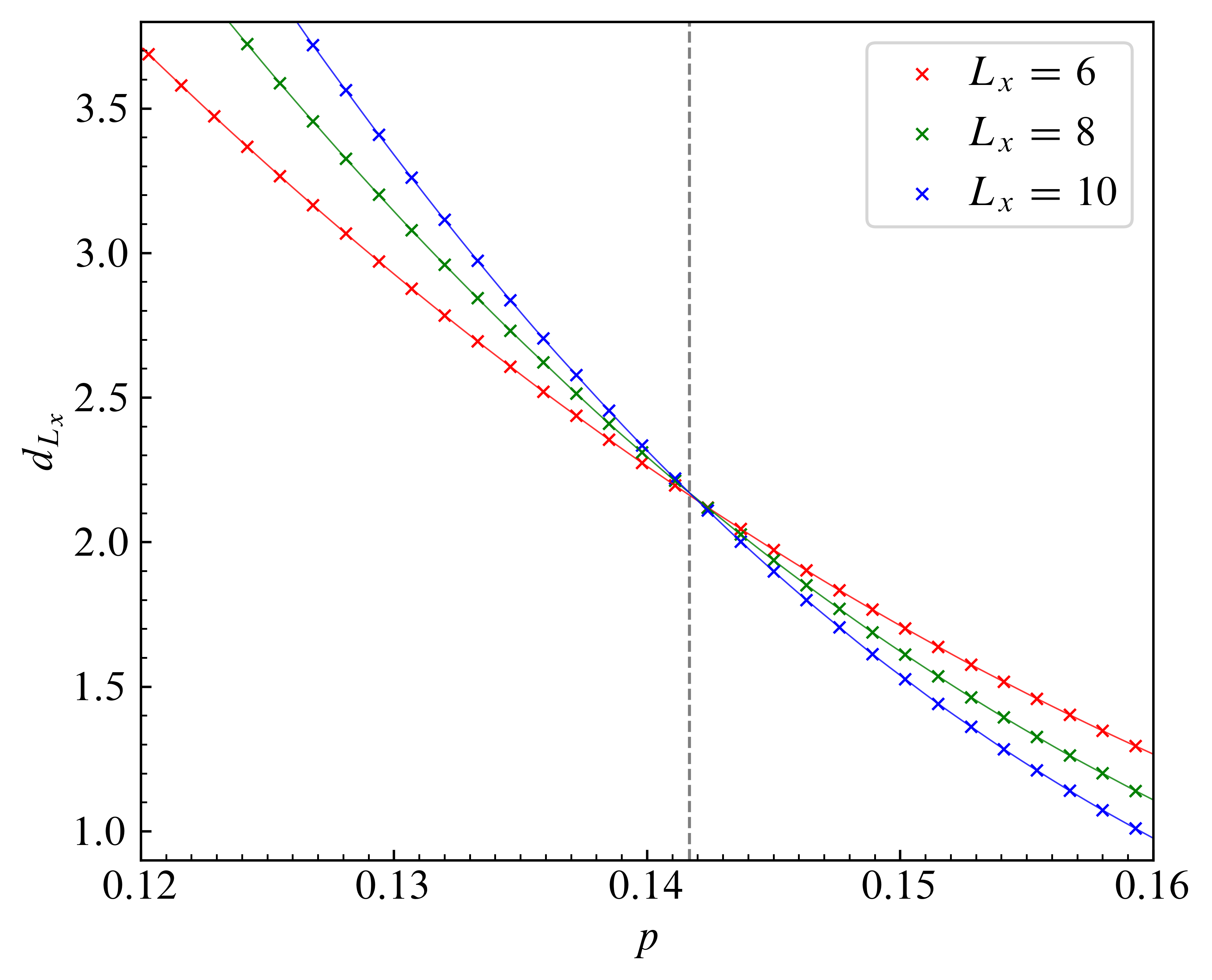}
\caption{\label{fig:plot_dw} The DW free energy of $L_x$-by-80000 lattices  of the SM model \cref{eq:SM_Hamiltonian}. }
\end{figure}

As the equilibration time of the parallel tempering MCMC simulation becomes overwhelmingly long when the phase boundary gets closer to the Nishimori surface, we switch to the numerical transfer-matrix method and use the DW free energy \cite{Honecker2001} to locate the Nishimori point. In this way, we consider an $L_x$-by-$L_y$ lattice with a free boundary in the $y$ direction; then the DW free energy is
\begin{align}
d_{L_x} = L_x^2 \beta \left(\mathbb{E}(f_{L_x}^\text{(ap)}) - \mathbb{E}(f_{L_x}^\text{(p)}) \right),
\end{align}
where $f_{L_x}^\text{(ap)}$ and $f_{L_x}^\text{(p)}$ are the free energies per site under antiperiodic and periodic boundary conditions in the $x$ direction, respectively. The $y$ direction should be as long as possible given the available computational resources so that the exponential growth rate of the partition function and thus $f_{L_x}$ can be obtained more accurately. In our work, $L_y=80000$ while $L_x = 6, 8, 10$, and each lattice size is averaged over 3960 disorder realizations. The small value of $L_x$ is due to the fact that the number of possible spin states in a row is $2^{L_x}$. This exponential scaling results in a roughly 22-hour run on a GPU for $L_x=10$ and 396 disorder realizations. See \cref{fig:plot_dw} for a sample plot.

Near the critical point, $d_{L_x}$ is expected to scale as
\begin{align}
d_{L_x} (p) = \mathcal{D} \left((p-p_c) L_x^{1/\nu}\right),
\end{align}
for some scaling function $\mathcal{D}$. Since our values of $L_x$ are not in pairs with a constant ratio, we choose to use the finite-size scaling analysis \verb"fssa" \cite{fssa} to find the best estimate of $p_c$. The idea is that the plot of $d_{L_x}$ against $(p-p_c) L_x^{1/\nu}$ as a whole should be the same for every $L_x$ since it is simply the plot of the function $\mathcal{D}$ itself; then \verb"fssa" uses the technique of data collapsing to find the best values of $p_c$ and $\nu$ so that the curves of different $L_x$ collapse into one.

For the zero-temperature point, in the case of the random-bond Ising model, the minimum-weight perfect matching algorithm was invented to find the ground state \cite{Picco2006}. However, since our model involves the four-spin plaquette interactions, we modify the DW free energy at finite temperature by considering the ground-state energy shift due to a DW. In this situation, we instead consider an $L$-by-$L$ square lattice with a free boundary in the $y$ direction and compute the zero-temperature DW energy \cite{Amoruso2004},
\begin{align}
d_L = E_L^\text{(ap)} - E_L^\text{(p)},
\end{align}
where $E^\text{(ap)}_L$ and $E^\text{(p)}_L$ are the ground-state energies under antiperiodic and periodic boundary conditions in the $x$ direction, respectively; then we compute 
\begin{align}
r_L = \frac{\mathbb{E}(d_L)^2}{\mathbb{E}(d_L^2)}.
\end{align}
$r_L$ is a dimensionless quantity, and we expect it to scale in a similar way near the critical point:
\begin{align}
r_L (p) = \mathcal{R} \left((p-p_c) L_x^{1/\nu}\right).
\end{align}
Based on our experience, $r_L$ is very sensitive to disorder fluctuations, and thus we have to use 79200 disorder realizations to smooth the plot of $r_L$ against $p$.

The cross sections of the phase diagram at $p=q$ and at $q=0.04$ are given in \cref{fig:phasediagram_pqsame,fig:phasediagram_q04}. In particular, if $p=0$ and $q\neq 0$, then $K=0$, and the SM model \cref{eq:SM_Hamiltonian} becomes the purely ferromagnetic Ising model, which has the famous critical temperature $2/\ln (1+\sqrt{2})$.

For decoding plaquette syndromes, since the SM model \cref{eq:SM_Hamiltonian_dual} involves two species, it is difficult to define the DW. Therefore, we choose to use the MCMC algorithm to estimate a possible range of $p_N$. See \cref{fig:phasediagram_pqsame,fig:phasediagram_q04}. A dedicated numerical method could be invented to locate the point more precisely. In particular, if $p=0$ and $q\neq 0$, then $K=0$, and the model becomes a collection of $N$ independent pairs, which has no ordered phase at finite temperature.

\section{The special case of perfect correctability}
In the absence of $Z$ errors, i.e., $q=0$ and $0<p<0.5$, we can show that the two SM models are in the ferromagnetic phase at any nonzero temperature and thus the QEC code is always correctable; that is, the error threshold is 0.5.

\subsection{Vertex syndromes}
For the correction of vertex syndromes, we consider the SM model \cref{eq:SM_Hamiltonian}. Since $q=0$ and $0<p<0.5$, $K=+\infty$, so the system is surely in the ground space of the four-spin interaction term at any temperature; otherwise, it would require an infinite amount of energy to be in an excited state. The ground states of the four-spin interaction term are the states where $-1$ spins form whole longitude and latitude lines so that the DWs do not have corners. This means the ground space is equivalent to a one-dimensional random-bond Ising model. Mathematically, $\sigma_{x,y}\sigma_{x+1,y}\sigma_{x,y+1}\sigma_{x+1,y+1}=1$ for all $x$, $y$ induces the decomposition
\begin{align}
\sigma_{x,y} = s_{x,1} s_{1,y}.
\end{align}
The nearest-neighbor interaction becomes $\sigma_{x,y}\sigma_{x+1,y} = s_{x,1} s_{x+1,1}$ in rows and $s_{1,y} s_{1,y+1}$ in columns, and the Hamiltonian restricted to this ground space is
\begin{align}
H = &-\sum_{x,y} \eta_{(x,y)(x+1,y)} s_{x,1} s_{x+1,1} \nonumber \\
&- \sum_{x,y} \eta_{(x,y)(x,y+1)} s_{1,y} s_{1,y+1}.\label{eq:SM_1dising_0}
\end{align}
For the corresponding QEC model, a pair of $h$ particles can only be caused by the two complementary horizontal $X$ strings ending at them, of which the union is a whole latitude in that row. Together with the $v$ particles, this confirms that the deformation of error configurations indeed makes the QEC model equivalent to the SM model whose possible configurations are made up of whole longitude and latitude DWs. Because of this fact, we can actually adopt the free boundary condition on an $L$-by-$L$ lattice. Let
\begin{subequations}
\begin{align}
J_x = \sum_{y=1}^L \eta_{(x,y)(x+1,y)}\, ,  \\
J_y = \sum_{x=1}^L \eta_{(x,y)(x,y+1)}\, ;
\end{align}
\end{subequations}
then the Hamiltonian \cref{eq:SM_1dising_0} is rewritten as
\begin{align}
H = - \sum_x J_x s_{x,1} s_{x+1,1} - \sum_y J_y s_{1,y} s_{1,y+1}. \label{eq:SM_1dising_1}
\end{align}
Further, since $s_{1,1}$ is shared by both terms, \cref{eq:SM_1dising_1} is essentially equivalent to a one-dimensional random-bond Ising model of $N=2L-1$ spins. Therefore, we can effectively reduce the Hamiltonian to
\begin{align}
H = - \sum_{i=1}^{N} J_i s_i s_{i+1} \label{eq:SM_1dising}
\end{align}
with 
\begin{align}
J_i = \sum_{j=1}^{N-1} \eta_j. \label{eq:SM_1dising_coupling}
\end{align}

The one-dimensional random-bond Ising model can be solved exactly. At inverse temperature $\beta$, the partition function is
\begin{align}
Z_J = \tr (T_1 ... T_N) ,
\end{align}
where every transfer matrix can be diagonalized by the same matrix
\begin{align}
P = \frac{1}{\sqrt{2}} \begin{pmatrix}1 & 1 \\ 1 & -1\end{pmatrix},
\end{align}
that is,
\begin{align}
T_i &= \begin{pmatrix} e^{\beta J_i} & e^{-\beta J_i} \\ e^{-\beta J_i} & e^{\beta J_i} \end{pmatrix}
= P D_i P, \\
D_i &= \begin{pmatrix} \lambda_i^+  & 0 \\ 0 & \lambda_i^- \end{pmatrix}
\end{align}
with 
\begin{subequations}
\begin{align}
\lambda_i^+ &= 2\cosh \beta J_i, \\
\lambda_i^- &= 2\sinh\beta J_i.
\end{align}
\end{subequations}
So, 
\begin{align}
Z_J = \lambda_1^+ ...\lambda^+_N \ +\ \lambda_1^- ...\lambda^-_N.
\end{align}
Assuming $j>i$, the two-spin correlation function is
\newpage\vspace*{-2.5\baselineskip}
\begin{align}
& \langle s_i s_j \rangle \nonumber \\
&= \frac{1}{Z_J} \tr (T_1 ... T_{i-1} Z_i  T_i ... T_{j-1} Z_j T_j ... T_N) \nonumber \\
&= \frac{1}{Z_J} \tr (P D_1 ... D_{i-1} P Z_i P D_i ... D_{j-1} P Z_j P D_j ... D_N P) \nonumber \\
&= \frac{1}{Z_J} \tr (X_i D_i ... D_{j-1} X_i D_j ... D_N D_1 ... D_{i-1} ),
\end{align}
where the classical $\mathbb{Z}_2$ variables are identified as the Pauli $Z$ matrices, and $PZ_iP=X$, the Pauli $X$ matrix. Explicitly,
\begin{align}
\langle s_i s_j \rangle 
=\,& \frac{\lambda_1^+ ...\lambda^+_N}{\lambda_1^+ ...\lambda^+_N + \lambda_1^- ...\lambda^-_N} \ 
\frac{\lambda^-_i ...\lambda^-_{j-1}}{\lambda^+_i ...\lambda^+_{j-1}} \nonumber \\
&+\  \frac{\lambda_1^- ...\lambda^-_N}{\lambda_1^+ ...\lambda^+_N + \lambda_1^- ...\lambda^-_N} \ 
\frac{\lambda^+_i ...\lambda^+_{j-1}}{\lambda^-_i ...\lambda^-_{j-1}}.
\end{align}
In the thermodynamic limit, $N\rightarrow\infty$, since $\lambda^+_i > \lambda_i^-$,
\begin{align*}
\frac{\lambda_1^+ ...\lambda^+_N}{\lambda_1^+ ...\lambda^+_N + \lambda_1^- ...\lambda^-_N} &\rightarrow 1, \\
\frac{\lambda_1^- ...\lambda^-_N}{\lambda_1^+ ...\lambda^+_N + \lambda_1^- ...\lambda^-_N} &\rightarrow 0.
\end{align*}
Thus, 
\begin{align}
\langle s_i s_j \rangle =
\frac{\lambda^-_i ...\lambda^-_{j-1}}{\lambda^+_i ...\lambda^+_{j-1}}
= \prod_{k=i}^{j-1} \tanh (\beta J_k).
\end{align}
Since $J_k$ are independent, the disorder average is
\begin{align}
\mathbb{E}(\langle s_i s_j \rangle) &= \mathbb{E}\big(\tanh(\beta J)\big)^{j-i}.
\end{align}
By \cref{eq:SM_1dising_coupling} and the law of large numbers, $J/(N-1) \rightarrow \mathbb{E}(\eta) = 1(1-p) + (-1)p = 1-2p > 0$ as $N\rightarrow\infty$. Then, $J\rightarrow\infty$ and $\tanh(\beta J)\rightarrow 1$ as $N\rightarrow\infty$. Also, $|\tanh(\beta J)| \leqslant 1$ for every system size $N$. By the dominated convergence theorem, $\lim_{N\rightarrow\infty}\mathbb{E} (\tanh(\beta J) ) = \mathbb{E} (\lim_{N\rightarrow\infty}\tanh(\beta J) ) = \mathbb{E}(1) = 1$; that is, $\mathbb{E}(\langle s_i s_j \rangle)=1$ as $N\rightarrow\infty$. Thus, by \cref{eq:magnet2}, the magnetization squared $\mathbb{E}(\langle M^2\rangle) = 1$ in the thermodynamic limit. The system is in the ferromagnetic phase at any temperature, in particular on the Nishimori surface.

\subsection{Plaquette syndromes}
For the correction of plaquette syndromes, we consider the SM model \cref{eq:SM_Hamiltonian_dual}. Since $q=0$, each layer is a collection of independent one-dimensional Ising models. For $0<p<0.5$, $K=+\infty$, so each layer must be in its own ground space: all $\tau_{x,y}$ having the same $x$ are the same, and all $\sigma_{x,y}$ having the same $y$ are the same. Let us call
\begin{align}
\begin{aligned}
\tau_x = \tau_{x,y} \quad \forall y=1, ..., L; \\
\sigma_y = \sigma_{x,y} \quad \forall x = 1, ..., L.
\end{aligned}
\end{align}
The Hamiltonian \cref{eq:SM_Hamiltonian_dual} becomes
\begin{align}
H = -\sum_{x=1}^L \sum_{y=1}^L \eta_{x,y} \tau_x \sigma_y.
\end{align}
It contains interactions between every $\tau_x$ and $\sigma_y$ and thus has no exact solution so far. Here we present a mean-field argument. 

Let us write 
\begin{align}
\eta_{x,y} = \bar\eta + \delta \eta_{x,y},
\end{align}
where
\begin{align}
& \mathbb{E}(\eta_{x,y}) = \bar\eta = 1-2p, \quad \text{Var}(\eta_{x,y}) = 1-\bar\eta^2, \\
& \mathbb{E}(\delta\eta_{x,y}) = 0. \nonumber
\end{align}
A $\sigma_y$ spin feels a field of
\begin{align}
\sum_x \eta_{x,y} \tau_x = \bar\eta \sum_x \tau_x + \sum_x \delta \eta_{x,y} \tau_x . \label{eq:meanfield}
\end{align}
The disordered term in \cref{eq:meanfield} has zero expectation and the variance $L(1-\bar\eta^2)$, so its typical value is about $\sqrt{L}$, but the first term in \cref{eq:meanfield} scales with $L$ and thus dominates in the thermodynamic limit, $L\rightarrow\infty$. Let $m_\sigma = \langle \sigma_y\rangle$ and $m_\tau = \langle \tau_x\rangle$. So, under the mean-field approximation, a typical disorder realization gives a field of $\bar\eta \sum_x \tau_x \simeq \bar\eta m_\tau L$. On the other hand, a $\mathbb{Z}_2$ spin in a field $h$ has a thermal expectation $\tanh(\beta h)$. Thus, the mean-field equation is
\begin{align}
m_\sigma = \tanh (\beta\bar\eta m_\tau L).
\end{align}
The same argument also gives $m_\tau = \tanh (\beta\bar\eta m_\sigma L)$. Since $\sigma_x$ and $\tau_x$ do not differ and are thus symmetric in our model, $m_\sigma = m_\tau = m$ and
\begin{align}
m = \tanh (\beta\bar\eta m L).
\end{align}
To find $T_c$, expand the equation near $m=0$, and the leading order of the solution is $\beta_c \bar\eta L = 1$. Since $\bar\eta = 1-2p>0$, this tells us that $T_c \propto L$. So, $T_c\rightarrow\infty$ as $L\rightarrow\infty$; the model is in the ferromagnetic phase at any temperature.

\section{Peierls's arguments for the error thresholds}
In addition to the numerical results, we provide an analytical lower-bound estimate of the Nishimori point of the SM model \cref{eq:SM_Hamiltonian} by using a disordered version \cite{Roepstorff} of Peierls's argument \cite{Peierls1936,Griffiths1964,Bonati2014}. 

\subsection{Disorder-free case ($p=q=0$)}
We first consider the disorder-free version of the model \cref{eq:SM_Hamiltonian}, that is, the case of $p=q=0$. The Hamiltonian then has a $\mathbb{Z}_2$ symmetry, $\sigma_i \mapsto -\sigma_i$, and the magnetization (per spin)
\begin{align}
M = \frac{1}{N} \sum_i \sigma_i = 1 - \frac{2N_-}{N}
\end{align}
has zero thermal average, where $N$ is the total number of spins and $N_-$ is the total number of $-1$ spins. In order to break the symmetry and study the spontaneous magnetization, we set all spins on the boundary to $+1$. Then the ground state is the state with all spins being $+1$. The DW, which is also called Peierls's contour, is defined as a loop on the dual lattice that separates the antialigned nearest neighbors, and a DW configuration is in one-to-one correspondence with a spin configuration because of the $+1$ boundary condition. The nearest-neighbor interaction counts the length of the DWs, while the four-spin plaquette interaction counts the corners: if $\sigma_i \sigma_j \sigma_k \sigma_l=-1$, the DW must have a 90-degree corner at the center of that plaquette. Therefore, a DW is uniquely determined by the length $L$, the number of corners $C$, and an index $i$ that is assigned at will to distinguish the DWs with the same $L$ and $C$. We shall write a DW as $\gamma_{L,C,i}$. Their possible values are
\begin{subequations}
\begin{align}
L &= 4, 6, 8, \, \ldots; \\
C &= 4, 6, 8,\, \ldots, L; \\
i &= 1, 2, 3,\, \ldots, \#(L, C),
\end{align}
\end{subequations}
where $\#(L, C)$ is the number of DWs of length $L$ and $C$ corners. A DW can be treated as an oriented path on the dual lattice, and thus every corner is either a left or a right turn. Let the number of left-turn corners be $C_L$ and the number of right-turn corners be $C_R$, then $C = C_L + C_R$. For the DW to be a closed loop, $|C_L - C_R| = 4$. We choose to label the DW such that $C_L > C_R$, that is, a counterclockwise loop. So, $C = 2C_L - 4$. The number of DWs with length $L$ and $C$ corners is
\begin{align*}
{L \choose C_L} {L-C_L \choose C_R} = {L \choose \frac{C}{2}+2} {L-\frac{C}{2}-2 \choose \frac{C}{2}-2}.
\end{align*}
Even if we knew specifically which turn occurs at which edge, we still could not uniquely determine a DW because the first edge can still be chosen freely. For a $\sqrt{N}\times\sqrt{N}$ lattice, the first edge can be any one of the $2N - 2\sqrt{N}$ bonds, and we shall simply use $2N$ for the upper bound. In fact, we only know $L$ and $C_L$. So, a DW can be obtained by starting from any of its $L$ edges. Thus, the number of DWs is bounded by
\begin{align}
\#(L, C) \leqslant \frac{1}{L} \cdot 2N \cdot {L \choose \frac{C}{2}+2} {L-\frac{C}{2}-2 \choose \frac{C}{2}-2}. \label{eq:2disingcorner_number}
\end{align}
Further, any DW can be contained in a rectangle, and thus the area enclosed is $\leqslant (L/4)^2$. The largest area is obtained when $C=4$, that is a rectangle, while increasing $C$ by 2 decreases the enclosed area by 1 or even more. Therefore,
\begin{align}
\text{Area}(\gamma_{L, C, i}) \leqslant \left(\frac{L}{4}\right)^2 - \frac{C-4}{2} 
= \frac{L^2}{16} -\frac{C}{2}+ 2 .\label{eq:2disingcorner_area}
\end{align}
Because of the $+1$ boundary condition and the fact that a DW can be inside another DW, a $-1$ spin must be contained in at least one DW. So, the number of $-1$ spins must be $\leqslant$ the total area of all DWs present:
\begin{align}
N_- \leqslant \sum_L \sum_C \sum_i \text{Area}(\gamma_{L,C,i}) \Theta(\gamma_{L,C,i}) ,\label{eq:2disingcorner_num_avg_0}
\end{align}
where $\Theta(\gamma_{L, C, i}) = 1$ if $\gamma_{L, C, i}$ is present in the configuration and $\Theta(\gamma_{L, C, i})=0$ if it is absent. Taking the thermal average on both sides and using \cref{eq:2disingcorner_area}, we have
\begin{align}
\langle N_-\rangle \leqslant \sum_L\sum_C \left(\frac{L^2}{16} -\frac{C}{2}+ 2\right) \sum_i \langle\Theta(\gamma_{L, C, i})\rangle. \label{eq:2disingcorner_num_avg}
\end{align}
The thermal average $\langle\Theta(\gamma_{L, C, i})\rangle$ is given by
\begin{align}
\langle\Theta(\gamma_{L, C, i})\rangle = \frac{\displaystyle\sum_{s} \Theta(\gamma_{L, C, i}) e^{-\beta E(s)}}{\displaystyle\sum_{s} e^{-\beta E(s)}}
= \frac{\displaystyle\sum_{g} e^{-\beta E(g)}}{\displaystyle\sum_{s} e^{-\beta E(s)}} ,\label{eq:2disingcorner_indicator_avg}
\end{align}
where $s$ is a configuration that satisfies the $+1$ boundary condition, and $g$ is a configuration that satisfies the $+1$ boundary condition and contains $\gamma_{L, C, i}$. Now consider the configuration $g'$ that is obtained from $g$ by flipping all the spins inside $\gamma_{L, C, i}$, then the DW $\gamma_{L, C, i}$ disappears in $g'$, and the energy 
\begin{align}
E(g') = E(g) - 2L - 2KC,
\end{align} 
but $g'$ still satisfies the $+1$ boundary condition, so $\{ g'\}$ is a subset of $\{s\}$, that is,
\begin{align*}
\sum_{g'} e^{-\beta E(g')} \leqslant \sum_{s} e^{-\beta E(s)}.
\end{align*}
Then by \cref{eq:2disingcorner_indicator_avg}, 
\begin{align}
\langle\Theta(\gamma_{L, i})\rangle \leqslant
e^{-2\beta L - 2\beta KC},
\end{align}
where we used the fact that the number of configurations $g$ is the same as the number of configurations $g'$. Plug this back into \cref{eq:2disingcorner_num_avg} and use \cref{eq:2disingcorner_number},
\begin{align*}
\langle N_-\rangle 
\leqslant \sum_L\sum_C &\left(\frac{L^2}{16} -\frac{C}{2}+ 2\right) \sum_i e^{-2\beta L - 2\beta KC} \\
\leqslant \sum_L\sum_C &\left(\frac{L^2}{16} -\frac{C}{2}+ 2\right) \\
&\times\frac{2N}{L} {L \choose \frac{C}{2}+2} {L-\frac{C}{2}-2 \choose \frac{C}{2}-2} e^{-2\beta L - 2\beta KC}.
\end{align*}
Let 
\begin{align}
x = e^{-2\beta},  \qquad y = e^{-2\beta K} = x^K;\label{eq:2disingcorner_Peierls_ineq_xy}
\end{align}
this becomes
\begin{align}
\frac{\langle N_-\rangle }{N} 
\leqslant\ 
\frac{1}{8}&\sum_L L x^L \sum_C {L \choose \frac{C}{2}+2} {L-\frac{C}{2}-2 \choose \frac{C}{2}-2} y^C  \nonumber\\
-&\sum_L \frac{1}{L} x^L \sum_C C {L \choose \frac{C}{2}+2} {L-\frac{C}{2}-2 \choose \frac{C}{2}-2} y^C  \nonumber\\
+ 4&\sum_L \frac{1}{L} x^L \sum_C {L \choose \frac{C}{2}+2} {L-\frac{C}{2}-2 \choose \frac{C}{2}-2} y^C \nonumber\\
\equiv I(x,y) & . 
\label{eq:2disingcorner_Peierls_ineq}
\end{align}
All these sums can be computed exactly. Let
\begin{align}
\Delta_\pm = \sqrt{(1\pm x)^2 - 4x^2 y^2},
\end{align}
then the first term in $I(x,y)$ is
\begin{subequations}\label{eq:2disingcorner_Peierls_ineq_I}
\begin{align}
x \frac{\partial}{\partial x} \left( x^4 y^4\left(\frac{1}{\Delta_-(1-x+\Delta_-)^4}+\frac{1}{\Delta_+(1+x+\Delta_+)^4}\right) \right),
\end{align}
the third term is
\begin{align}
8 x^4 y^4\left(\frac{1}{(1-x+\Delta_-)^4}+\frac{1}{(1+x+\Delta_+)^4}\right),
\end{align}
and the second term is simply $-\frac{1}{4}\, y \frac{\partial}{\partial y}$ of the third term, so it is
\begin{align}
-2 y \frac{\partial}{\partial y} \left(x^4 y^4\left(\frac{1}{(1-x+\Delta_-)^4}+\frac{1}{(1+x+\Delta_+)^4}\right)\right).
\end{align}
\end{subequations}
We analyze this upper bound in three cases. 
\begin{itemize}
\item At $x=0$ and $y=0$, that is, $T\rightarrow 0$, the upper-bound function $I(x,y)$ is zero, so $\langle N_-\rangle = 0$; that is, the magnetization is $+1$. This simply confirms that ``all $+1$'' is the unique ground state. In the vicinity of $x=0$ and $y=0$, the upper-bound function is less than $1/2$, that is, $\frac{\langle N_-\rangle}{N} < \frac{1}{2}$ and thus $\langle M\rangle > 0$. This proves the existence of a ferromagnetic phase at sufficiently low $T$.  
\item For $K=1$, that is, $y=x$, demanding that the upper-bound function $I(x,y)<\frac{1}{2}$ leads to $0 < x < 0.4757$ and thus $0 < k_\mathrm{B} T < 2.6921$. This is the minimal region of the ferromagnetic phase. Indeed, our numerical result shows $k_\mathrm{B} T_c = 3.3053$.
\item For $K=0.2$, that is, $y=x^{0.2}$, demanding that the upper-bound function $I(x,y)<\frac{1}{2}$ leads to  $0 < x < 0.3547$ and thus $0< k_\mathrm{B} T < 1.9243$. The numerical simulation \cite{Fu2017} gives $k_\mathrm{B} T_c= 2.493$.
\end{itemize}

\newpage
\subsection{General case ($0<p,q<0.5$)}
Now consider the general model with disorder. Even with disordered bonds, the DW is still defined as a loop on the dual lattice that separates the antialigned nearest neighbors. The number of $-1$ spins is still $\leqslant$ the area of all DWs, and \cref{eq:2disingcorner_num_avg_0,eq:2disingcorner_num_avg,eq:2disingcorner_indicator_avg}
still hold. However, when we consider the configuration $g'$ that is obtained from $g$ by flipping all the spins inside $\gamma_{L,C,i}$, the energy relation is now different:
\begin{align}
E(g') = E(g) - 2J (L - 2\ell) -2K(C-2c),
\end{align}
where $\ell$ is the number of edges of $\gamma_{L, C, i}$ going through an antiferromagnetic bond and $c$ is the number of corners residing in a disordered plaquette. Since $g'$ still has the $+1$ boundary, $\{g\} \cup \{ g'\}$ is a subset of $\{ s\}$, that is,
\newpage
\begin{align*}
\sum_{g} e^{-\beta E(g)} + \sum_{g'} e^{-\beta E(g')} \leqslant \sum_{s} e^{-\beta E(s)} 
\end{align*}
and 
\begin{align}
\langle\Theta(\gamma_{L, C, i})\rangle 
&\leqslant \frac{1}{1 + e^{2\beta J(L-2\ell)+2\beta K(C-2c)}}.
\end{align}
Since the probability of having $\ell$ out of $L$ edges going through the antiferromagnetic bonds and $c$ out of $C$ corners residing in the disordered plaquettes is
\begin{align*}
{L \choose \ell} p^\ell (1-p)^{L-\ell} {C \choose c} q^c (1-q)^{C-c},
\end{align*}
the disordered average is
\begin{widetext}
\begin{align}
\mathbb{E}\langle\Theta(\gamma_{L, C, i})\rangle
\leqslant \sum_{\ell=0}^L \sum_{c=0}^C 
& {L \choose \ell} p^\ell (1-p)^{L-\ell} {C \choose c} q^c (1-q)^{C-c} \frac{1}{1 + e^{2\beta J(L-2\ell)+2\beta K(C-2c)}}.
\end{align}
Using the identity
\begin{align}
{L \choose \ell} = \frac{1}{\pi} \int_0^\pi \rmd u\ (2\cos u)^L \cos(L-2\ell)u \label{eq:Peierls_identity}
\end{align}
for both ${L \choose \ell}$ and ${C \choose c}$, and letting $-2m = L-2\ell$ and $-2n=C-2c$, we have
\begin{align}
\mathbb{E}\langle\Theta(\gamma_{L, C,i})\rangle \leqslant 
\frac{1}{\pi^2} \int_0^\pi \rmd u &\int_0^\pi\rmd v \ \phi(u, v) 
\left(2\sqrt{p(1-p)}\cos u\right)^L 
\left(2\sqrt{q(1-q)}\cos v\right)^C ,
\end{align}
where
\begin{align}
\phi(u,v) = \sum_{m,n=-\infty}^{\infty} 
\left(\frac{p}{1-p}\right)^m \left(\frac{q}{1-q}\right)^n  
\frac{1}{1+e^{-4\beta Jm-4\beta Kn}} \cos (2mu) \cos (2nv).
\end{align}
Strictly speaking, $m=-L/2, \ldots, L/2$ and $n=-C/2, \ldots, C/2$, but, since $0<p,q<0.5$, very large $m$ and $n$ make the term extremely small, so we can extend the sums to infinity. In this way, $f$ is independent of $L$ and $C$. Since the model stays on the Nishimori surface, \cref{eq:Nishimori} reduces it to
\begin{align}
\phi(u,v) 
&= \sum_{m,n=-\infty}^{\infty}  \frac{e^{-2\beta m -2\beta Kn} }{1+e^{-4\beta m-4\beta Kn}} \cos (2mu) \cos (2nv) \nonumber \\
&= \sum_{m,n=-\infty}^{\infty}  \frac{1}{2}\, \mathrm{sech} (2\beta m+2\beta K n) \cos (2mu) \cos (2nv)  \nonumber \\
&= \frac{\pi^2}{8\beta} \sum_{j,k\in\mathbb{Z}}  \mathrm{sech}\left(\frac{\pi}{2\beta}(j \pi-u)\right)  \Big(\delta(K(j \pi-u)+v-k\pi) + \delta(K(j \pi-u)-v-k\pi)\Big), \label{eq:Peierls_phi}
\end{align}
where we used the Poisson summation formula of the sech function (see Appendix \ref{app:Peierls_calculation}). Let 
\begin{subequations}
\begin{align}
x(u) &= 2\sqrt{p(1-p)}\cos u ,\\
y(v) &= 2\sqrt{q(1-q)}\cos v.
\label{eq:2disingcorner_disorder_Peierls_ineq_xy}
\end{align}
\end{subequations}
Then by \cref{eq:2disingcorner_num_avg,eq:2disingcorner_number},
\begin{align}
\frac{\mathbb{E}\langle N_-\rangle}{N} 
&\leqslant \sum_L \sum_C 
\left(\frac{L^2}{16} -\frac{C}{2}+ 2\right) 
\frac{2}{L} {L \choose \frac{C}{2}+2} {L-\frac{C}{2}-2 \choose \frac{C}{2}-2}\, \mathbb{E}\langle\Theta(\gamma_{L, C, i})\rangle \nonumber\\
& \leqslant \frac{1}{\pi^2} \int_0^\pi \rmd u \int_0^\pi\rmd v\ \phi(u, v)\, I (x(u),y(v)) ,\label{eq:2disingcorner_disorder_Peierls_ineq}
\end{align}
where $I (x(u),y(v))$ takes exactly the same form as \cref{eq:2disingcorner_Peierls_ineq}, of which the explicit result is \cref{eq:2disingcorner_Peierls_ineq_I}. 

Finally, let us specialize to the case of $p=q$, then $K=1$ and  
\begin{align}
\phi(u,v) = \frac{\pi^2}{8\beta} \sum_{j\in\mathbb{Z}} 
\mathrm{sech}\left(\frac{\pi}{2\beta} (u-j\pi)\right) (D_\pi(u-v) + D_\pi(u+v)), \label{eq:Peierls_phi_K1}
\end{align}
where $D_\pi$ is a Dirac comb at $\mathbb{Z}\pi$, that is,
\begin{align}
D_\pi (u) \equiv \sum_{n\in\mathbb{Z}} \delta(x-n\pi) . \label{eq:Dirac_comb}
\end{align}
The sum of sech over $j$ can be written using the elliptic theta functions (see Appendix \ref{app:Peierls_calculation}):
\begin{align}
\phi(u,v) = \frac{\pi}{4} \vartheta_3\left(0,\frac{p}{1-p}\right)\vartheta_4\left(0,\frac{p}{1-p}\right) \frac{\vartheta_3\left(v,\frac{p}{1-p}\right)}{\vartheta_4\left(v,\frac{p}{1-p}\right)} (D_\pi(u-v) + D_\pi(u+v)), \label{eq:Peierls_phi_K1_elliptic}
\end{align}
where we used the definition of the Nishimori surface \cref{eq:Nishimori} and the elliptic theta functions
\begin{subequations}\label{eq:elliptic_theta}
\begin{align}
\vartheta_3(z,r) &= \sum_{n\in\mathbb{Z}} r^{n^2} e^{2\mathrm{i}nz}, \\
\vartheta_4(z,r) &= \sum_{n\in\mathbb{Z}} (-1)^n r^{n^2} e^{2\mathrm{i}nz}. 
\end{align}
\end{subequations}
For the integration on $0<u<\pi$ and $0<v<\pi$, $\phi(u,v)$ is nonzero only on the two diagonal lines $v=u$ and $v=\pi-u$, so
\begin{align}
& \frac{1}{\pi^2} \int_0^\pi \rmd u \int_0^\pi\rmd v\ \phi(u, v)\, I(x(u),y(v)) \nonumber\\
=\,& \frac{1}{4\pi} \vartheta_3\left(0,\frac{p}{1-p}\right)\,\vartheta_4\left(0,\frac{p}{1-p}\right) \int_0^\pi \rmd u\, \frac{\vartheta_3\left(u,\frac{p}{1-p}\right)}{\vartheta_4\left(u,\frac{p}{1-p}\right)} \Big( I(x(u),y(u)) + I(x(u),y(\pi-u)) \Big) \nonumber \\
=\,& \frac{1}{2\pi} \vartheta_3\left(0,\frac{p}{1-p}\right)\,\vartheta_4\left(0,\frac{p}{1-p}\right) \int_0^\pi \rmd u\, \frac{\vartheta_3\left(u,\frac{p}{1-p}\right)}{\vartheta_4\left(u,\frac{p}{1-p}\right)} \, I(x(u),x(u)) , \label{eq:2disingcorner_disorder_samepq}
\end{align}
\end{widetext}
where we used $y(\pi-u) = - y(u)$, $y(u)=x(u)$ for $p=q$, and $I(x,-y) = I(x,y)$. This upper bound, \cref{eq:2disingcorner_disorder_samepq}, decreases to 0 as $p\rightarrow 0$, which means the Nishimori surface goes into the ferromagnetic phase at sufficiently low $p$. Demanding \cref{eq:2disingcorner_disorder_samepq} $<\frac{1}{2}$ gives $p < 0.0654$. Indeed, our numerical result, \cref{eq:p_N}, is larger than this. A similar Peierls's argument applied to the random-bond Ising model gives $0.0269$ for the lower bound on $p_\mathrm{N}$ (see Appendix \ref{app:Peierls_random_bond_ising}), which supports the conclusion that the new QEC code indeed increases $p_\mathrm{N}$ for vertex syndromes.

\section{Conclusions and outlook}
We studied the error thresholds of a new, symmetry-enriched toric code under independent Pauli $X$ and $Z$ errors. The decoding problems of vertex syndromes and of plaquette syndromes were mapped to two different classical disordered SM models. We used the parallel tempering Monte Carlo simulation and the numerical transfer matrix algorithm to explore the phase diagrams of these models. A disordered version of Peierls's argument was provided for a lower-bound estimate of the error threshold of the vertex syndromes. We also showed that this QEC code has perfect correctability if Pauli $Z$ errors are absent. 

There are still many directions and topics to explore. For the SM models, we can study whether the four-spin interaction gives rise to the spin-glass phase at nonzero temperature, as the two-dimensional random-bond Ising model only has the spin-glass phase above $p_0$ at zero temperature \cite{Honecker2001,Picco2006,Amoruso2004}. By improving the numerical methods, we could compare $p_0$ to $p_N$ more precisely and confirm the reentrance of the ferromagnetic phase, as the numerical results \cite{Picco2006,Amoruso2004} of the two-dimensional random-bond Ising model suggest reentrance when $p_0 < p < p_N$. For the QEC code, we can further include the measurement errors, which will turn the model into a three-dimensional lattice in spacetime. We could also consider the depolarizing noisy channel, 
\begin{align}
\rho \ \mapsto\ (1-r)\rho + \frac{r}{3}X \rho X + \frac{r}{3} Y\rho Y + \frac{r}{3} Z\rho Z.
\end{align}
In this channel, $X$ and $Z$ are correlated, yet the toric code has been found to have a higher error threshold \cite{Bombin2012}. Therefore, we conjecture a higher error threshold for our model under this channel. Another way to increase the error threshold is to change the geometry of the lattice since certain geometries, such as the triangular lattice, allow a higher error threshold for the toric code \cite{Geometry}. 

Our preliminary investigation also shows that the threshold for plaquette syndromes may be smaller than that of the toric code provided that the Pauli $X$ and $Z$ errors are equally probable. This suggests studying a variant of our model in which the plaquette syndromes also have associated corner excitations. Although there is no self-duality, this decrease in the plaquette threshold, together with the increase in the vertex threshold, is reminiscent of \cite{english2026dualityconstrainsoptimalthresholds}, where the authors showed that in self-dual models, there is an inverse relationship between the thresholds of dual excitation types, with the square-lattice toric code being optimal.

\begin{acknowledgments}
This work used computational and storage services associated with the Hoffman2 Cluster which is operated by the UCLA Office of Advanced Research Computing’s Research Technology Group. The numerical programs are written by J.H. in the C language and only translated into the CUDA language with the help of Codex. GPT 5.6 Sol was used to find the special-function identities in \cite{NIST:DLMF} that are used in Appendix \ref{app:Peierls_calculation}. R.T. is supported by the Mani L. Bhaumik Presidential Term Chair. R.T. is grateful to Takamasa Ando and Shuhei Ohyama for collaboration on our related work \cite{ohyama2026parameterizedfamiliestoriccode} where we introduced the symmetry-enriched toric code we studied here. R.T. is also grateful to Ruben Verresen for suggesting to study the error correction properties of it.
\end{acknowledgments}

\appendix
\newpage
\appsection{Duality of the Two SM Models}
\label{app:duality}
Let us call $\eta_{(x,y)(x,y+1)} = \lambda_{x,y}$ and $\eta_{(x,y)(x+1,y)} = \rho_{x,y}$, then \cref{eq:SM_Hamiltonian_dual} is rewritten as
\begin{align}
H_\eta =& -\sum_{x,y} \eta_{x,y} \sigma_{x,y} \tau_{x,y} \nonumber \\
&- K \sum_{x,y} \lambda_{x,y} \tau_{x,y} \tau_{x, y+1} \nonumber\\
&- K\sum_{x,y} \rho_{x,y}\sigma_{x,y} \sigma_{x+1, y}. \label{eq:duality_H}
\end{align}
At inverse temperature $\beta$, the partition function is 
\begin{align}
Z_\eta = \sum_{\{ \sigma, \tau\}} \prod_{x,y} 
&\ e^{\beta \eta_{x,y}\sigma_{x,y}\tau_{x,y}} \nonumber \\[-4mm]
& \times e^{\beta K\lambda_{x,y}\tau_{x,y}\tau_{x,y+1}} \nonumber \\
& \times e^{\beta K\rho_{x,y}\sigma_{x,y}\sigma_{x+1,y}} .
\end{align}
Since $\eta_{x,y}=\pm 1$ and $\sigma_{x,y}\tau_{x,y}=\pm 1$, we have the following identity:
\begin{align}
&e^{\beta \eta_{x,y}\sigma_{x,y}\tau_{x,y}} \nonumber \\
=\,& A(\beta) \sum_{c_{x,y}=\pm 1} 
e^{\beta^* c_{x,y}} (\eta_{x,y} \sigma_{x,y}\tau_{x,y})^{ (1-c_{x,y})/2},
\end{align}
where
\begin{align}
e^{-2\beta^*} = \tanh \beta \label{eq:duality_1}
\end{align}
and 
\begin{align}
A(\beta) = \sqrt{\frac{\sinh(2\beta)}{2}}.
\end{align}
Similarly, we also have
\begin{align}
& e^{\beta K\lambda_{x,y}\tau_{x,y}\tau_{x,y+1}} \nonumber\\
=\,& A(\beta K) \sum_{b_{x,y}=\pm 1} e^{\beta^* K^* b_{x,y}} (\lambda_{x,y} \tau_{x,y}\tau_{x,y+1})^{ (1-b_{x,y})/2}, \\
& e^{\beta K\rho_{x,y}\sigma_{x,y}\sigma_{x,y+1}} \nonumber\\
=\,& A(\beta K) \sum_{a_{x,y}=\pm 1} e^{\beta^* K^* a_{x,y}} (\rho_{x,y} \sigma_{x,y}\sigma_{x+1,y})^{ (1-a_{x,y})/2},
\end{align}
where 
\begin{align}
e^{-2\beta^* K^*} = \tanh \beta K. \label{eq:duality_2}
\end{align}
Note that, in \cref{eq:duality_1,eq:duality_2}, we choose to define the dual inverse temperature and the coupling constant so that $K=1$ corresponds to $K^*=1$. So, the partition function becomes
\newpage\vspace*{-2\baselineskip}
\begin{align*}
Z_\eta =\,& A(\beta K)^{2N^2} A(\beta)^{N^2} 
\sum_{\{ \sigma, \tau\}} \sum_{\{a\}} \sum_{\{b\}}\sum_{\{c\}} \prod_{x,y} \\
& \hphantom{\times~} e^{\beta^* c_{x,y}} (\eta_{x,y} \sigma_{x,y}\tau_{x,y})^{ (1-c_{x,y})/2} \\
& \times e^{\beta^* K^* b_{x,y}} (\lambda_{x,y} \tau_{x,y}\tau_{x,y+1})^{ (1-b_{x,y})/2} \\
& \times e^{\beta^* K^* a_{x,y}} (\rho_{x,y} \sigma_{x,y}\sigma_{x+1,y})^{ (1-a_{x,y})/2} \\[2mm]
=\,& A(\beta K)^{2N^2} A(\beta)^{N^2} \sum_{\{ \sigma, \tau\}} \sum_{\{a\}} \sum_{\{b\}}\sum_{\{c\}} \\
& e^{\beta^*\sum_{x,y} c_{x,y} + K^* (a_{x,y} + b_{x,y})} \\
& \prod_{x,y}\ (\eta_{x,y} \sigma_{x,y}\tau_{x,y})^{ (1-c_{x,y})/2} \\[-2mm]
& \ \times (\lambda_{x,y} \tau_{x,y}\tau_{x,y+1})^{ (1-b_{x,y})/2} \\
& \ \times (\rho_{x,y} \sigma_{x,y}\sigma_{x+1,y})^{ (1-a_{x,y})/2}.
\end{align*}
Now let us consider the sum over $\sigma$ configurations. The exponent of $\sigma_{x,y}$ is $(1-c_{x,y})/2 + (1-a_{x-1,y})/2 + (1-a_{x,y})/2$. Since $\sum_{\sigma=\pm 1} \sigma^n =2$ for even $n$ and $=0$ for odd $n$, only the terms with $
c_{x,y} a_{x-1,y} a_{x,y} = 1$ survive, that is, 
\begin{align}
c_{x,y} = a_{x-1,y} a_{x,y}.
\end{align}
In the same way, for the sum over $\tau$ configurations, only the terms with 
\begin{align}
c_{x,y} = b_{x,y-1} b_{x,y}
\end{align}
survive. These two requirements induce the following substitution: 
\begin{align}
a_{x,y} &= S_{x+\frac{1}{2}, y-\frac{1}{2}} S_{x+\frac{1}{2}, y+\frac{1}{2}}, \\
b_{x,y} &= S_{x-\frac{1}{2}, y+\frac{1}{2}} S_{x+\frac{1}{2}, y+\frac{1}{2}},
\end{align}
where $\{ S\}$ are $\pm 1$ variables defined on the vertices of the dual lattice. Then,
\begin{align}
c_{x,y} = S_{x-\frac{1}{2}, y-\frac{1}{2}} S_{x-\frac{1}{2}, y+\frac{1}{2}} S_{x+\frac{1}{2}, y-\frac{1}{2}} S_{x+\frac{1}{2}, y+\frac{1}{2}},
\end{align}
which is the four-spin interaction around a plaquette of the dual lattice. The partition function becomes
\begin{align}
Z_\eta \nonumber \\
=\,& A(\beta K)^{2N^2} A(\beta)^{N^2}  \sum_{\{a\}} \sum_{\{b\}}\sum_{\{c\}} 
e^{\beta^*\sum_{x,y} c_{x,y} + K^* (a_{x,y} + b_{x,y})} \nonumber\\[-1mm]
& 2^{N^2} 2^{N^2} \prod_{x,y} 
\eta_{x,y}^{ (1-c_{x,y})/2} 
\lambda_{x,y}^{ (1-b_{x,y})/2}
\rho_{x,y}^{ (1-a_{x,y})/2}\Bigg|_{\begin{subarray}{l} c_{x,y} \\ = a_{x-1,y}a_{x,y} \\ = b_{x,y-1} b_{x,y}\end{subarray}} \nonumber\\ 
=\,& A(\beta K)^{2N^2} A(\beta)^{N^2} 
\sum_{\{S\}} e^{-\beta^* H^*}\nonumber \\
& 2^{N^2} 2^{N^2} \prod_{x,y} 
\eta_{x,y}^{ (1-S_{x-\frac{1}{2}, y-\frac{1}{2}} S_{x-\frac{1}{2}, y+\frac{1}{2}} S_{x+\frac{1}{2}, y-\frac{1}{2}} S_{x+\frac{1}{2}, y+\frac{1}{2}})/2}  \nonumber\\[-3mm]
& \qquad\qquad \times \lambda_{x,y}^{ (1-S_{x-\frac{1}{2}, y+\frac{1}{2}} S_{x+\frac{1}{2}, y+\frac{1}{2}})/2} \nonumber\\
& \qquad\qquad \times \rho_{x,y}^{ (1-S_{x+\frac{1}{2}, y-\frac{1}{2}} S_{x+\frac{1}{2}, y+\frac{1}{2}})/2} 
\label{eq:duality_partition}
\end{align}
with the Hamiltonian
\begin{align}
H^* = & -\sum_{x,y} S_{x-\frac{1}{2}, y-\frac{1}{2}} S_{x-\frac{1}{2}, y+\frac{1}{2}} S_{x+\frac{1}{2}, y-\frac{1}{2}} S_{x+\frac{1}{2}, y+\frac{1}{2}} \nonumber \\
&- K^* \left(S_{x+\frac{1}{2}, y-\frac{1}{2}} S_{x+\frac{1}{2}, y+\frac{1}{2}} + S_{x-\frac{1}{2}, y+\frac{1}{2}} S_{x+\frac{1}{2}, y+\frac{1}{2}}\right).
\end{align}
This is exactly the disorder-free version of the other model, \cref{eq:SM_Hamiltonian}. However, we cannot say that the partition functions of the two models are equivalent because of the extra product in \cref{eq:duality_partition}. Only in the case of no disorder, that is, $\eta_{x,y}=\lambda_{x,y}=\rho_{x,y}=1$,  does the product over $x$ and $y$ in \cref{eq:duality_partition} become 1 and
\begin{align}
Z = \text{const.} \times \sum_{\{S\}} e^{-\beta^* H^*}, 
\end{align}
where $Z$ now becomes the partition function of the disorder-free version of \cref{eq:duality_H}, that is,
\begin{align}
H = &-\sum_{x,y} \sigma_{x,y} \tau_{x,y} \nonumber \\
&- K \sum_{x,y} \tau_{x,y} \tau_{x, y+1} - K\sum_{x,y} \sigma_{x,y} \sigma_{x+1, y} .
\end{align}
Finally, the critical temperature of $H^*$ with $K^*=1$ is related to that of $H$ with $K=1$ by \cref{eq:duality_1}. Indeed, our numerical results confirm this: $k_\mathrm{B}T_c^* =1.6224$ and $k_\mathrm{B}T_c = 3.3053$. See \cref{fig:phasediagram_pqsame}.

\appsection{Calculation of the $\phi$ Function in Peierls's Argument}
\label{app:Peierls_calculation}
We present the details of the calculation of the $\phi$ function in Peierls's Argument, that is, \cref{eq:Peierls_phi_K1} and \cref{eq:Peierls_phi_K1,eq:Dirac_comb,eq:Peierls_phi_K1_elliptic}. Both require the Fourier transform of the sech function, 
\begin{align}
\int_{-\infty}^\infty \sech(ax) e^{-i\xi x} \rmd x = \frac{\pi}{a} \sech\left(\frac{\pi}{2a}\xi\right), \label{eq:FT_sech}
\end{align}
and the Poisson summation formula,
\begin{align}
\sum_{n\in\mathbb{Z}} f(a+bn) = \frac{1}{b} \sum_{m\in\mathbb{Z}} \tilde{f} \left( \frac{2\pi m}{b} \right) e^{i \frac{2\pi m}{b}a}, \label{eq:Poisson_sum}
\end{align}
where $\tilde{f}$ is the Fourier transform of $f$.

\subsection{Calculation of \cref{eq:Peierls_phi}}\vspace*{-4mm}
\begin{align}
\phi(u,v) \nonumber \\
= \sum_{m,n=-\infty}^{\infty} &\frac{1}{2}\, \sech (2\beta m+2\beta K n) \cos (2mu) \cos (2nv)  \nonumber \\
= \sum_{m,n=-\infty}^{\infty} &\frac{1}{2}\, \sech (2\beta (m+K n)) \times \frac{1}{4} \Big(e^{2i(mu+nv)} \nonumber \\[-1mm]
&+ e^{2i(mu-nv)} + e^{2i(-mu+nv)} + e^{2i(-mu-nv)}\Big). \nonumber
\end{align}
\newpage\noindent
Substituting $m$ and $n$ by $-m$ and $-n$ in the last two terms and using the fact that sech is an even function, we have
\begin{align}
\phi (u,v) &= \sum_{m,n=-\infty}^{\infty} \frac{1}{2}\, \sech (2\beta (m+K n))  \nonumber \\
&\qquad\qquad\times \frac{1}{2} \Big(e^{2i(mu+nv)} + e^{2i(mu-nv)} \Big) \nonumber \\
&= \frac{1}{2}(F(u,v) + F(u,-v)), \label{eq:Peierls_phi_FF}
\end{align}
where we call 
\begin{align}
F(u,v) \equiv \sum_{m,n=-\infty}^{\infty} \frac{1}{2}\, \sech (2\beta (m+K n)) e^{2i(mu+nv)} .
\end{align}
By the inverse Fourier transform of sech, 
\begin{align*}
&F(u,v) \\
&= \sum_{m,n} \frac{1}{4\beta} \int_{-\infty}^\infty \rmd \xi \,\sech\left(\frac{\pi}{2\beta}\xi\right) e^{2i\xi(m+Kn)} e^{2i(mu+nv)}  \\
&= \frac{1}{4\beta} \int_{-\infty}^\infty \rmd \xi \,\sech\left(\frac{\pi}{2\beta}\xi\right) \sum_m e^{2im(\xi+u)} \sum_n e^{2in(\xi K+v)}.
\end{align*}
By \cref{eq:Poisson_sum}, $\pi \sum_{j\in\mathbb{Z}} \delta(x-j\pi) =  \sum_{n\in\mathbb{Z}} e^{2inx}$,
\begin{align}
&F(u,v) \nonumber \\
&= \frac{\pi^2}{4\beta} \sum_{j,k\in\mathbb{Z}} \int_{-\infty}^\infty \rmd \xi \,\sech\left(\frac{\pi}{2\beta}\xi\right) \nonumber \\
&\qquad\qquad\qquad \times \delta(\xi+u-j\pi)\, \delta(\xi K+v-k\pi) \nonumber \\
&= \frac{\pi^2}{4\beta} \sum_{j,k\in\mathbb{Z}} \sech\left(\frac{\pi}{2\beta} (j\pi-u)\right)\delta((j\pi-u)K+v-k\pi).
\end{align}
Plugging this back into \cref{eq:Peierls_phi_FF}, we get \cref{eq:Peierls_phi}.

\subsection{Calculation of \cref{eq:Peierls_phi_K1,eq:Dirac_comb,eq:Peierls_phi_K1_elliptic}}
By the Fourier transform \cref{eq:FT_sech}, 
\begin{align}
\int_{-\infty}^\infty \sech\left(\frac{\pi}{2\beta}x\right) e^{-i\xi x} \rmd x = 2\beta\, \sech( \beta\xi);
\end{align}
then by the Possion summation formula \cref{eq:Poisson_sum},
\begin{align}
&\sum_{j\in\mathbb{Z}} \sech \left(\frac{\pi}{2\beta} (u-j\pi) \right) \nonumber \\
=\,& \frac{1}{\pi} \sum_{n\in\mathbb{Z}} 2\beta\, \sech(2\beta n) e^{i2nu} \nonumber \\
=\,& \frac{2\beta}{\pi} \sum_{n\in\mathbb{Z}}  \sech(2\beta n) \cos(2nu) \nonumber \\
=\,& \frac{2\beta}{\pi} \left( 1+ 2 \sum_{n=1}^\infty \frac{2e^{-2\beta n}}{1 + e^{-4\beta n}} \cos(2nu) \right) .
\end{align}
\newpage\noindent
On the Nishimori surface \cref{eq:Nishimori}, 
\begin{align}
&\sum_{j\in\mathbb{Z}} \sech \left(\frac{\pi}{2\beta} (u-j\pi) \right) \nonumber \\
=\,& \frac{2\beta}{\pi} \left( 1+ 4 \sum_{n=1}^\infty \frac{\left(\frac{p}{1-p}\right)^n}{1 + \left(\frac{p}{1-p}\right)^{2n}} \cos(2nu) \right) .
\end{align}
The sum of this form can be found in \cite{NIST:DLMF}, namely Eq. (22.11.3) with the definitions Eqs. (22.2.2) and (22.2.6), that is,
\begin{align}
&\sum_{j\in\mathbb{Z}} \sech \left(\frac{\pi}{2\beta} (u-j\pi) \right) \nonumber \\
=\,& \frac{2\beta}{\pi} \vartheta_3\left(0,\frac{p}{1-p}\right)\vartheta_4\left(0,\frac{p}{1-p}\right) \frac{\vartheta_3\left(v,\frac{p}{1-p}\right)}{\vartheta_4\left(v,\frac{p}{1-p}\right)},
\end{align}
where $\vartheta_3$ and $\vartheta_4$ are defined in \cref{eq:elliptic_theta}. Plugging this back into \cref{eq:Peierls_phi_K1}, we can get \cref{eq:Peierls_phi_K1_elliptic}.

\appsection{Peierls's Argument for the Random-Bond Ising Model}
\label{app:Peierls_random_bond_ising}
For the random-bond Ising model, the first edge of a DW of length $L$ can be put on any one of the $2N - 2\sqrt{N}$ bonds, and we simply use $2N$ for the upper bound. For the 2nd, 3rd, ..., $(L-1)$th edges, it can either turn left, turn right, or keep straight, giving three choices. For the last edge, it has only one choice in order to close the loop. Therefore, the number of DWs of length $L$ is
\begin{align}
\# (L) \leqslant \frac{2N \cdot 3^{L-2}}{L} = \frac{2N}{9L} 3^{L}.\label{eq:2dising_number}
\end{align}
Since the DW is a loop on the square lattice, it has the maximum area when it is a square, so
\begin{align}
\text{Area} (\gamma_{L,i}) \leqslant \left(\frac{L}{4}\right)^2 = \frac{L^2}{16}.\label{eq:2dising_area}
\end{align}
Instead of \cref{eq:2disingcorner_num_avg,eq:2disingcorner_indicator_avg}, we have
\begin{align}
\langle N_-\rangle \leqslant \sum_L \frac{L^2}{16} \sum_i \langle\Theta(\gamma_{L, i})\rangle \label{eq:2dising_num_avg}
\end{align}
and
\begin{align}
\langle\Theta(\gamma_{L, i})\rangle = \frac{\displaystyle\sum_{g} e^{-\beta E(g)}}{\displaystyle\sum_{s} e^{-\beta E(s)}},
\end{align}
where $g$ is the configuration that satisfies the $+1$ boundary condition and contains the DW $\gamma_{L,i}$. Again, consider the configuration $g'$ that is obtained from $g$ by flipping all the spins inside $\gamma_{L,i}$, then the DW $\gamma_{L, i}$ disappears in the new configuration $g'$, and 
\begin{align}
E(g') = E(g) - 2J (L - 2\ell),
\end{align}
where $\ell$ is the number of edges going through an antiferromagnetic bond. Since $\{ g\}\cup\{ g'\}$ is a subset of $\{ s\}$,
$$\sum_{g} e^{-\beta E(g)} + \sum_{g'} e^{-\beta E(g')}\leqslant \sum_{s} e^{-\beta E(s)},$$
and hence
\begin{align}
\langle\Theta(\gamma_{L, i})\rangle &\leqslant \frac{1}{1 + e^{2\beta J(L-2\ell)}}, \\
\mathbb{E}\langle\Theta(\gamma_{L, i})\rangle &\leqslant \sum_{\ell=0}^L {L \choose \ell} p^\ell (1-p)^{L-\ell} \frac{1}{1 + e^{2\beta J(L-2\ell)}}.
\end{align}
Using the identity \cref{eq:Peierls_identity} and letting $-2m = L-2\ell$, we have
\begin{align}
\mathbb{E}\langle\Theta(\gamma_{L, i})\rangle \leqslant 
\frac{1}{\pi}\int_0^\pi \rmd t\,\phi(t) \left(2\sqrt{p(1-p)}\cos t\right)^L,
\end{align}
\newpage\noindent where, by extending the range of $m$ to infinity, 
\begin{align}
\phi(t) = \sum_{m=-\infty}^{\infty} \left(\frac{p}{1-p}\right)^m \frac{1}{1+e^{-4\beta Jm}} \cos 2mt.
\end{align}
Then, 
\begin{align*}
& \mathbb{E}(\langle N_-\rangle )\\
&\leqslant \sum_L \frac{L^2}{16} \sum_i \mathbb{E}\langle\Theta(\gamma_{L, i})\rangle \\
&\leqslant \sum_L \frac{L^2}{16} \frac{2N}{9L} 3^{L} \frac{1}{\pi} \int_0^\pi \rmd t\, \phi(t)  \left(2\sqrt{p(1-p)}\cos t\right)^L ,
\end{align*}
that is, 
\begin{align}
\frac{\mathbb{E}\langle N_-\rangle}{N} 
&\leqslant  \frac{1}{72\pi} \int_0^\pi \rmd t\, \phi(t)\, \bm{\mathsf{P}} (6\sqrt{p(1-p)}\cos t) ,
\end{align}
where we define 
\begin{align}
\bm{\mathsf{P}} (t) = \sum_{L=4,6, ...} Lt^L  = \frac{2t^4(2-t^2)}{(1-t^2)^2}.
\end{align}
Along the Nishimori line $e^{-2\beta} = p/(1-p)$,
\begin{align*}
\phi(t) = \frac{1}{2}\vartheta_3\left(0,\frac{p}{1-p}\right)\,\vartheta_4\left(0,\frac{p}{1-p}\right) \frac{\vartheta_3\left(t,\frac{p}{1-p}\right)}{\vartheta_4\left(t,\frac{p}{1-p}\right)}.
\end{align*}
Numerical integration shows that, if $\frac{\mathbb{E}\langle N_-\rangle}{N} < \frac{1}{2}$, then $p < 0.0269$. This is the lower bound on $p_\mathrm{N}$.

\nocite{*}
\bibliography{threshold}

\end{document}